\documentclass[iop, revtex4]{emulateapj-rtx4}

\shorttitle{Metal Abundances of 12 ADBS dwarf irregulars}
\shortauthors{Haurberg et al.}

\usepackage[parfill]{parskip}    
\usepackage{longtable}
\usepackage{epsfig}
\usepackage{graphicx}
\usepackage{amssymb}
\usepackage{epstopdf}
\usepackage{lscape}
\usepackage{natbib}
\newcommand{\simlt}{\ {\raise-.5ex\hbox{$\buildrel<\over\sim$}}\ }
\newcommand{\hi}{\ion{H}{1}}
\newcommand{\hii}{\ion{H}{2}}

\newcommand{\ha}{H$\alpha$}

\newcommand{\oiiite}{[\ion{O}{3}]$\,\lambda4363$}
\newcommand{\oiii}{[\ion{O}{3}]$\,\lambda \,\lambda4959, 5007$} 
\newcommand{\rtt}{([\ion{O}{3}]$\,\lambda \,\lambda4959,5007$+[\ion{O}{2}]]$\,\lambda3727$)/{H}$\beta$}
\newcommand{\ott}{[\ion{O}{3}]$\,\lambda \,\lambda4959,5007$/[\ion{O}{2}]$\,\lambda3727$}
\newcommand{\niioii}{[\ion{N}{2}$]\,\lambda6583$/[\ion{O}{2}]$\,\lambda3727$}
\newcommand{\te}{T$_{e}$}

\begin{document}
\bibliographystyle{plainnat}

\title{Metal Abundances of 12 Dwarf Irregulars from the ADBS Survey}

\author{Nathalie C. Haurberg\altaffilmark{1}, Jessica Rosenberg\altaffilmark{2}, \& John J. Salzer\altaffilmark{1}}

\altaffiltext{1}{Department of Astronomy, Indiana University, 727 E. Third St., Bloomington, IN 47405; nhaurber@astro.indiana.edu, slaz@astro.indiana.edu}
\altaffiltext{2}{School of Physics, Astronomy and Computational Science, George Mason University, MS 3F3, Fairfax, VA 22030.; jrosenb4@gmu.edu}
                                   
\pagebreak

\begin{abstract}

We have analyzed long-slit spectra of 12 dwarf irregular galaxies from the Arecibo Dual Beam Survey (ADBS).  These galaxies represent a heterogeneous sample of objects detected by ADBS, but on average are relatively gas-rich, low-surface-brightness, and low-mass, thus represent a region of the galaxian population that is not commonly included in optical surveys.  The metallicity-luminosity relationship for these galaxies is analyzed; the galaxies discussed in this paper appear to be under-abundant at a given luminosity when compared to a sample from the literature.  We attempt to identify a ``second parameter" responsible for the intrinsic scatter and apparent under-abundance of our galaxies.  We do not find a definitive second parameter but note the possible indication that infall or mixing of pristine gas may be responsible.  We derive oxygen abundances for multiple \hii\ regions in many of our galaxies but do not find any strong indications of metallicity variation within the galaxies except in one case where we see variation between an isolated \hii\ region and the rest of the galaxy. Our data set includes the galaxy with the largest known \hi\,-\,to\,-\,optical size ratio, ADBS 113845+2008.  Our abundance analysis of this galaxy reveals that it is strongly over-enriched compared to galaxies of similar luminosity, indicating it is not a young object and confirming the result from \citet{can1138} that this galaxy appears to be intrinsically rare in the local universe.  
\end{abstract}


\clearpage


\section{INTRODUCTION}
\label{sec:intro}

The Arecibo Dual Beam Survey (ADBS), completed in 1999  \citep{rs2000}, was one of the most sensitive and complete surveys for gas-rich galaxies in the local universe at the time of its undertaking.  The survey covered approximately 430 square degrees and detected 265 confirmed extragalactic sources. The final catalog included 81 sources not previously identified from optical catalogs, including 7 low \hi\ mass (M$_{\textrm{\hi}}$) galaxies with M$_{\textrm{\hi}} < 10^{8}$ M$_{\odot}$, nearly doubling the number known at that time.  The majority of the ADBS catalog was included in broadband optical and narrowband H$\alpha$ follow-up studies performed on the WIYN 0.9 m at Kitt Peak National Observatory \citep{stev_adbsI, sudg_adbs}.  

The primary science goals of the ADBS were to determine if there exists an overlooked population of galaxies that could be identified with a blind \hi\ survey (such as galaxies that are gas-rich but star-poor) and to better understand the nature of gas-rich galaxies by filling in the lower portion of the \hi\ mass function \citep{rs2002}.  These queries have continued to be examined by both the \hi\ Parkes All Sky Survey \citep[HIPASS: e.g.,][]{kil99, kil00, mey04, zwa04, doy05} and the ongoing and far more sensitive Arecibo Legacy Fast ALFA (ALFALFA) survey \citep[e.g.,][]{giov05, haynes11, cannon11}.  In addition to providing preliminary insight into these issues, the ADBS produced a unique catalog of potential star-forming galaxies in the local universe with many viable science applications.   

We seek to better understand the chemical evolution of low-luminosity and gas rich systems by determining and analyzing the nebular abundances of 12 galaxies selected from the ADBS catalog, including ADBS 113845+2008, the galaxy with the largest known \hi\,-\,to\,-\,optical size ratio \citep{can1138}.  A spectroscopic follow-up study of these 12 galaxies was carried out at the MMT observatory. We use high-quality spectra of actively star-forming regions within these systems to determine chemical abundances and investigate trends with other galaxy parameters. The galaxies in our sample all have narrowband H$\alpha$ imaging \citep{sudg_adbs}, and all but one have broadband optical imaging.  

It has been established by several authors that, in general, the metallicity of star-forming galaxies decreases with decreasing luminosity leading to the so called luminosity-metallicity (L-Z) relationship \citep[e.g.,][]{Leq79, skill89, pil01, trem04, salz05}.  This suggests that low-mass and low-luminosity galaxies should be among the most metal-poor systems known in the local universe.  Studies of the luminosity-metallicity relationship have shown that it is, perhaps, more fundamentally a mass-metallicity (M-Z) relationship where luminosity is often the observed parameter that corresponds most directly to mass \citep[e.g.,][]{lee06, salz05, trem04}.  If the relationship is truly one between mass and metallicity, it suggests that the gas recycling process in low-mass galaxies differs significantly from massive galaxies and thus presents intriguing questions concerning the efficiency of star formation and the role of enriched gas outflows and pristine gas inflows in these systems.    Additionally, some authors have identified low-mass and low-luminosity galaxies that appear to have relatively high abundances \citep{zah12, peep08}, suggesting that they deviate from the established relationships and that the relationships may not apply universally to low-mass systems.  

The galaxies in our data set are low-mass dwarf irregulars (dIrrs) that were selected from the ADBS catalog as those with the lowest \hi-masses and lowest luminosity optical counterparts.  Since these galaxies were chosen from a blind \hi\ survey, they form a unique portion of the low-mass population that is not biased by optical properties.  Any galaxy included in the survey must be relatively gas-rich and some studies have suggested gas content plays a crucial role in understanding why and how dIrrs fit into the M-Z and L-Z relationships and how these systems form and evolve over time \citep[e.g.,][]{mag12, zah12, greb03}.

The role of dIrrs in the larger context of galaxy formation also raises interesting questions.  It is possible that gas-rich dIrrs can passively evolve into gas-poor dwarf spheroidals (dSphs) as they exhaust or lose their gas \citep{greb03}.  This could help solve the well known problem that standard $\Lambda$CDM models predict more dSph's than are observed in the local universe \citep[e.g.,][]{kly99}.  For this scenario to hold, one would expect the most gas-rich dIrrs to also be the most underenriched.  We see some evidence of this trend in our sample, however argue this trend could arise as the result of the infall of pristine gas, similar to the argument presented by \citet{edm90} and invoked by \citet{dal04} to explain metallicity trends in spiral galaxies.

We describe our sample in detail in Section 2, and our observations and reduction process in Section 3.  In Section 4 we present our spectral data, Section 5 covers the abundance determination, and Section 6 our abundance analysis.  The discussion of our results is in Section 7, including a sub-section devoted to ADBS 113845+2008 (Section 7.3), and our conclusions are in Section 8.

\section{DESCRIPTION OF SAMPLE}
\label{sec:samp}

The sample used for this project consists of 12 \hi-selected galaxies from the ADBS survey \citep{rs2000}.  The ADBS was a ``blind" \hi\ survey which provided a comprehensive catalog of gas-rich galaxies that is not biased by optical properties such as surface brightness and luminosity. Galaxies analyzed in this paper were additionally included in the optical and \ha\ follow-up studies of \citet[][optical broadband]{stev_adbsI} and \citet[][\ha]{sudg_adbs}, providing us a set of accurate magnitudes, $B - V$ colors, surface brightnesses, and \ha\ fluxes.  The one exception is ADBS 071352+1031 which was not included in the broadband study but has been previously analyzed by several other authors \citep{mak99, tul06, geo08, PS08, sha08} allowing us to include it in our sample by using photometric properties from the literature.

The galaxies in our sample were chosen based on their low \hi\ mass, which identified them as likely dIrrs.  With one exception, the distance to our galaxies all fall between 30 and 60 Mpc and have recessional velocities between 1700 and 4000 km s$^{-1}$.  We additionally required that each galaxy in the sample have high signal-to-noise \ha\ images so that individual \hii\ regions could be identified and selected for spectroscopic observations; examples of the \ha\ images are included in Figure \ref{fig:finder}.  Only a few target \hii\ regions were chosen for each galaxy, however, since we performed long-slit spectroscopy, we also obtained spectra for several other \hii\ regions that happened to fall in the slit when it was placed over the target regions.

Basic parameters for each of the galaxies in our sample were calculated and are listed in Table \ref{tab:basic}. For three of the galaxies (ADBS 071352+1031, 125156+1205, and 135822+2533) we have used a velocity-independent distance indicator from the source given in the table.  For the remaining galaxies the redshift values from \citet{rs2000} were used to derive distances based on the peculiar velocity flow models of \citet{mas05}, the same model that is used to determine  distances for galaxies with $cz$ $<$ 6000 km s$^{-1}$ in the ALFALFA $\alpha$.40 catalog \citep{haynes11}.  Distance determination for galaxies in the local universe is notoriously difficult and the uncertainties associated with distances are difficult to characterize \citep[e.g.,][]{mas04}. For this work, we have assumed a distance error of 10\% for all distances and propagated this through to other quantities where appropriate (such as the error on the absolute B magnitude). 

Also included in Table \ref{tab:basic} is the observed B-band magnitude (m$_{\textrm{B,obs}}$), optical color (B $-$ V), central B-band surface brightness ($\mu_{\textrm{o,B}}$) which are taken from \citet{stev_adbsI}, except where otherwise noted, and the Galactic absorption corrections taken from \citet{sch98}.  The error given in M$_{\textrm{B}}$ accounts for both the photometric error and the error in the distance discussed above. The \hi\ mass was calculated using the integrated \hi\ flux from \citet{rs2000} assuming distance cited in Table \ref{tab:basic}.  The star-formation rate (SFR) was similarly derived using the \ha\ fluxes (corrected for both Galactic and internal absorption) calculated in \citet{sudg_adbs} and assuming the distance used in this work. The conversion relation between the \ha\ luminosity and SFR from \citet{ken98} is adopted [SFR (M$_{\odot}$ yr $^{-1}$) $=$ L$_{H\alpha}$/ (1.26 $\times$ 10$^{41}$ ergs s$^{-1}$)]. 

\section{OBSERVATIONS \& REDUCTION}
\label{sec:obs}

\subsection{Observations}
\label{subsec:obs}

All observations presented in this paper were carried out using the MMT\footnote{Observations reported here were obtained at the MMT Observatory, a joint facility of the Smithsonian Institution and the University of Arizona.} with the Blue Channel spectrograph.   Data acquisition occurred on three nights: 5 February 2006 and 28-29 May 2006.  We employed a 300 line mm$^{-1}$ grating blazed at 4800 \AA\ that yielded a reciprocal dispersion of 1.95 \AA\ pixel$^{-1}$ on the CCD detector.   The total spectral coverage was 3600 -- 8500 \AA.  All sources were observed through a 1.5\arcsec\ wide slit, and the slit extended 180\arcsec\ in the spatial direction.

Each source was visually centered in the slit, since the slit-viewing camera reached deep enough to show our targets.   The position angle of the slit was adjusted to be close to the parallactic angle.  This was done to eliminate problems caused by differential atmospheric refraction.  In several cases it was possible to rotate the slit slightly off of the parallactic angle and include an additional HII region.   For several sources a number of additional faint HII regions fell within our slit serendipitously and were observed.  The final slit positions are illustrated in Figure 1.

We obtained spectra of several spectrophotometric standard stars each night, selected from the list of \citet{mass88}.  These were used to calibrate the flux scale in our spectra.  Additional calibration data included zero exposure time bias images, HeNeAR lamp spectra for wavelength calibrations, internal quartz lamp spectra for flat-field corrections, and spectra of the twilight sky that were used to correct for variations in the width of the slit.

\subsection{Data Reduction}
\label{subsec:data}

The two-dimensional (2D) spectra were processed through the standard spectral reduction routines in \texttt{IRAF}.  The overscan region was used to determine the bias level for each image. Median combined dome flats were used to correct for pixel-to-pixel variations and twilight flats were used to create an illumination function that accounts for large-scale 2D structures in the flat field.  Bad pixels and columns were interpolated over and all of the science images were then processed through the   \texttt{lacos\_spec} cosmic ray rejection routine of \citet{vd01}.  Each cosmic ray cleaned image was examined by eye and compared to the original image to ensure that emission line pixels were not rejected by the program.  

The spectra of individual \hii\ regions were carefully extracted from the spectral images.  Great care was taken in this process because, in many cases, the slit was placed through the central region of the galaxy resulting in several \hii\ regions as well as copious amounts of extended continuum emission from the galaxy falling in the slit.  Some of these \hii\ regions were packed tightly together on the image, leading to blended emission lines. Strong lines were examined carefully to choose appropriate apertures for extraction.  The extended continuum emission complicated background sky measurements as well, so we exercised similar attention when choosing the sky regions and evaluating the results of sky subtraction.  When possible we selected ``local" sky regions that closely bracketed our target emission regions, but when too much underlying galaxy light was present we opted for  more distant sky regions that avoided continuum from the underlying stellar population. The details of the line identification process are included in Section 3.3.

The wavelength scale for each spectrum was determined using the wavelength solution derived from HeNeAr lamp spectra. Spectrophotometric standard stars were then used to create a sensitivity function that was applied to all the spectra.  Once the sensitivity function was applied, each standard star was compared to a catalog spectrum for that star in order to check its validity.  Finally, the spectra were corrected for telluric absorption in each galaxy. 

\subsection{Identification of Individual \hii\ Regions}
\label{subsec:ident}

Due to the presence of multiple \hii\ regions along the slit in many of our 2D images, each region of \ha\ emission was examined along the dispersion axis to determine whether it appeared to be a legitimate emission-line source that could produce usable spectra.  We extracted all emission-line regions that appeared to contain several measurable lines (most notably \ha, H$\beta$, the [\ion{O}{2}] doublet, and the strong lines from the [\ion{O}{3}] triplet).   Once extracted, a combination of the 2D and one-deminsional (1D) spectra were used to match each \hii\ region with the bright knots of emission in the \ha\ images.  

To match the \hii\ regions accurately with the bright knots in Figure \ref{fig:finder} we began with the 2D images, which typically contain a well-centered target \hii\ region.  We then used the position angle of the slit to estimate where it cut through the galaxy. From the 2D spectra, which almost always contained more than one bright \hii\ region, we were able to better determine exactly how the slit intersected the galaxy.  In some cases it was still unclear, so we had to examine the 1D spectra to evaluate the relative strengths of \ha\ and continuum emission and compare this with the relative brightness of each knot in the \ha\ and R-band images.  After carefully going through several iterations of this process for each galaxy were able to identify the location within the narrowband \ha\ image of each \hii\ region we extracted.  In some cases one \hii\ region was partially in the slit in more than one image, so we compared the different spectra and opted to use whichever one contained stronger emission lines.  Once this process was completed we were able to create a finder chart for each galaxy (see Figure \ref{fig:finder}).  

We additionally identified two high-redshift background galaxies and one background quasar that serendipitously fell in the slit. A list of these objects along with their measured redshift is given in Table \ref{tab:hiz}.

\section{SPECTRAL DATA}
\label{sec:spec}

Figures \ref{fig:specfirst} \& \ref{fig:speclast} show examples of the final spectra analyzed for this project. In many cases, the quality of the spectra was such that even weak lines, such as \oiiite, were discernible.  In total, we extracted 16 unique \hii\ regions that contained a measurable \oiiite\ line; 4 of these are shown in Figure \ref{fig:specfirst}, where each spectrum contains a small arrow marking the location of this line. Figure \ref{fig:speclast} displays four examples of lower signal-to-noise spectra, where the weak lines were not detected. For each \hii\ region in Figures \ref{fig:specfirst} and \ref{fig:speclast} we have plotted two versions of the spectrum, one showing the full dynamic range of the emission and the other only a small range in flux so the weak lines can be seen. 

Emission lines in each spectrum were identified and measured using the \texttt{SPLOT} routine in \texttt{IRAF}.  We exercised considerable effort to identify all emission lines that were present, including weak lines. We first identified  strong lines and then searched for weak lines by carefully comparing each spectrum with a possible line-list.  When measuring line-fluxes, we set the continuum level using the average of the local continuum on both sides of the line. We used a linear fit for the continuum under the emission line and assumed that the continuum remained flat in the area under the line except in the case of very broad lines or regions where the continuum was particularly steep.

We expect the measurement of permitted emission lines (H-Balmer and He) from star-forming nebulae to be significantly affected by absorption lines in hot main-sequence stars (spectral types O B A).  To account for this absorption, we used line-ratios from multiple Balmer lines (H$\alpha$/H$\beta$, H$\gamma$/H$\beta$, and H$\delta$/H$\beta$) to simultaneously calculate the underlying absorption correction and the absorption coefficient at H$\beta$ (c(H$\beta$)).  The correct coefficient was chosen by determining at what underlying absorption the c({H$\beta$) coefficients matched for all the Balmer ratios. This process requires the assumption that the equivalent width of the underlying absorption lines is the same for all the Balmer lines.

Once the absorption corrections were applied to the hydrogen lines, the relative intensities of the Balmer lines are used to solve for the reddening using the reddening law of \citet{car89} and temperatures derived from the [\ion{O}{3}] lines when available. (If the \oiiite\ line was not detected, we assumed a electron temperature of 10,000 K.)  Since we use the Balmer decrement to calculate the reddening correction, we do not distinguish between intrinsic and foreground reddening for the spectra.  The values for the equivalent width of underlying absorption (W$_{\textrm{abs}}$) and c(H$\beta$) are included in Table \ref{tab:gold}. In some cases the Balmer-line data were not sufficient to obtain a reliable estimate of the underlying absorption-line equivalent width, such as cases where H$\alpha$ and H$\beta$ were the only available Balmer lines.  In these instances, we used an average underlying absorption value of $2 \AA$. We chose to use this average correction in lieu of no correction, because most \hii\ regions showed some significant amount of absorption and we felt it was necessary to attempt to account for this.  

The final results from our spectral analysis are tabulated in Table \ref{tab:gold}, along with W$_{\textrm{abs}}$, c(H$\beta$), \te, and n$_{e}$.   The electron temperature (T$_{e}$) and electron density (n$_{e}$) of the nebula were estimated using the \oiiite$/$\oiii\ and [\ion{S}{2}]$\,\lambda \,\lambda6716, 6731$ ratios, respectively.  For those galaxies with a measurable \oiiite\ line, and thus an estimate of \te, the temperature and density calculations were performed using the Emission Line Spectrum Analyzer (ELSA) program \citep{elsa} described in more detail in Section \ref{sec:te}. If a reliable detection of [\ion{S}{2}]$\,\lambda \,\lambda6716, 6731$ was not available we used a default electron density of 100 cm$^{-3}$, which is broadly consistent to densities commonly observed for \hii\ regions.  The name of each \hii\ region corresponds to its identification in Figure \ref{fig:finder}. In Table \ref{tab:gold} we have listed the reddening corrected flux relative to H$\beta$ for each line measured. The errors on the line fluxes are propagated fully through all calculation using the derived errors from the relevant quantities (e.g., rms of the continuum, rms scatter in flux calibration, etc.).

\section{ABUNDANCE DETERMINATION}
\label{sec:abun_det}

\subsection{Diagnostic Diagram}
\label{subsec:diag}

Figure \ref{fig:diag} shows an emission-line diagnostic diagram \citep[e.g.,][]{BPT, vo87} with each \hii\ region plotted separately.  The solid line in Figure \ref{fig:diag} is a theoretical curve from the models of \citet{de86} tracing the locus of high-excitation star-forming regions. The dashed line, empirically determined in \citet{kau03}, separates the region in the diagnostic diagram typically occupied by narrow-line active galactic nuclei (AGNs) from the region where stellar photoionized sources lie.

As expected, all our points appear to the left of the division between AGN and star-forming objects, indicating that they are emission nebulae powered by the ionizing radiation from O and B stars.  The points that fall significantly to the left of the model \hii\ region curve likely represent lower-excitation \hii\ regions where the hottest O stars have already left the main-sequence, resulting in a softer radiation field.  Most of our targeted \hii\ regions (plotted within diamonds) and most \hii\ regions with detectable \oiiite\ (plotted as filled points) sit fairly near the model curve as expected, since these should be some of the youngest and highest-excitation objects in our sample.

Figure \ref{fig:diag} can be used as a rough diagnostic for the metallicity of an \hii\ region.  The models of \citet{de86} vary smoothly in metallicity with the most metal-poor star-forming regions residing on the upper left portion and metallicity steadily increasing as you move down and to the right along the model curve.  Thus, star-forming galaxies with \hii\ regions near the upper left are generally considered metal-poor (i.e., dIrrs) and those with \hii\ regions that populate the lower right are more metal-rich (i.e., spirals).

There are two galaxies plotted in Figure \ref{fig:diag} that deserve special attention. The points in Figure \ref{fig:diag} that are plotted as \texttt{x}'s and stretch well into the ``metal-rich spiral" region of the diagram represent \hii\ regions in the galaxy  ADBS 125156+1205. This object was chosen for the project based on its apparently low \hi\ mass and late-type morphology, however an upward adjustment in its distance estimate has led to a factor of three increase in its \hi\ mass.  Additionally, its morphological classification was originally thought to be a Magellanic dwarf, but it is now clear that ADBS 125156+1205 is a more massive edge-on spiral.  Hence, this object does not fit in well with the rest of our sample and its properties will be evaluated separately throughout this paper. 

The other noteworthy galaxy is ADBS 113845+2008, represented by the triangle in Figure \ref{fig:diag}.   ADBS 113845+2008 is one of the most optically compact isolated galaxies currently known. Follow-up Very Large Array \hi\ mapping \citep{can1138}, revealed an enormous \hi\ disk that extends out to 44 B-band scale lengths around this source.  The nature of this galaxy is not yet fully understood, but its optically compact nature and abnormally extended \hi\ disk indicate that it has an unusual history. Therefore, we will consider it separately from the majority of our sample as well.  We will continue to use \texttt{x}'s to denote ADBS 125156+1205 and a triangle to denote ADBS 113845+2008 in subsequent plots.

\subsection{Methods of Abundance Determination}
\label{subsec:meth}

\subsubsection{The ``Direct" \te\ Method}
\label{sec:te}
The primary goal of this project has been to obtain accurate chemical abundances for the galaxies in this sample.  Elemental abundances of \hii\ regions can be determined using the relative line strengths of the nebular emission lines. However, since the relationship between line-strength and abundance is very temperature sensitive, an accurate estimate for the electron temperature (\te) of the nebula is necessary to determine an accurate abundance.  A direct \te\ measurement can be made using the relative strength of the temperature sensitive \oiiite\ line compared to the strength of the \oiii\ lines.  All of these lines are produced by the same ionization state of oxygen but the \oiiite\ and \oiii\ lines are produced by different energy level transitions; the rate of excitation to the different energy levels is highly sensitive to temperature and therefore the flux ratio of the corresponding emission lines is also temperature sensitive and can be used to measure the \te\ for the \hii\ region.  

The \oiiite\ line typically is weak and only measurable for \hii\ regions with moderate to high values of \te\ (because the strength of the \oiiite\ line decreases rapidly with decreasing temperature). We obtained spectra that allowed direct \te\ measurements for 16 of the 48 \hii\ regions analyzed in this study.  To calculate abundances using this ``direct" method (henceforth the \te\ method) we used the ELSA program described in detail in \citet{elsa}. \texttt{ELSA} was originally written for analyzing planetary nebula abundances but has been modified for use with star-forming galaxies.

The ELSA program uses a five-level atom routine with ionization correction factors and a two-region ionization model to calculate the elemental abundances.  The five-level atom calculation is based on that used by \citet{hen89} but has been significantly updated and improved from that original code \cite[see ][]{elsa}. The temperature and density for the high-ionization region are calculated from the standard \oiiite/\oiii\ and  [\ion{S}{2}]$\,\lambda \,\lambda6716, 6731$ ratios and an iterative procedure similar to that in \citet{izo06}. The electron temperature for the high-ionization region of a given nebula is what will be referred to as \te\ throughout this paper. The temperature for the low-ionization region typically is calculated from the [\ion{N}{2}]$\lambda$5755/[\ion{N}{2}]$\lambda\lambda$6548,6583  ratio.  In our case the ionization state is high enough that [\ion{N}{2}]$\lambda$5755 is not detectable, and the low-ionization region temperature is estimated using T$_{\textrm{e,OIII}}$ following the method of \citet{pag92}. Ionization correction factors are calculated using abundance ratios for measurable lines with similar ionization potentials and then multiplied to account for the unseen ionization states. The uncertainties in line fluxes, reddening correction, plasma diagnostics, and ionic abundances are propagated through the ELSA program and included in tabulating the final uncertainties.  

We also recognize that there may be unknown temperature fluctuations present, as noted in \citet{pei67}. These are most likely to bias us toward assuming a higher \te\ and thus lower metallicity.  It impossible to determine if these fluctuations are present, since the \hii\ regions we are examining are unresolved. If they are present, they may account for some of the scatter between abundances measured by the direct method and those determined by the ``McGaugh method" outlined in Section 5.2.2.

Abundance results for \hii\ regions with a measurable \oiiite\ line, and therefore \te\ abundance, are presented in Table \ref{tab:elsa}.  Column 1 lists the \hii\ region as identified in Figure \ref{fig:finder} and Column 2 lists the oxygen abundance tabulated as log(O/H) $+$ 12.  Columns 4 - 8  give abundance ratios relative to hydrogen for several key elements (He, N, O, Ne, S, and Ar).  All the errors cited are the errors calculated by the ELSA program as described above.

\subsubsection{The Strong-line McGaugh Method}
Many galaxies in our sample did not have an unambiguous detection of \oiiite, so we additionally calculated the oxygen abundance for each \hii\ region using the strong-line method described in McGaugh (1991; henceforth the McGaugh method).  To calculate abundances via the McGaugh method the following strong-line ratios are utilized: 
\begin{description}
\item[R$_{23}$] \rtt
\item[O$_{23}$] \ott
\item[ [\ion{N}{2}\textrm{]}/[\ion{O}{2}\textrm{]} ] \niioii
\end{description}
Using the R$_{23}$ and O$_{23}$ values, each \hii\ region is plotted on a grid of theoretical models (Figure \ref{fig:mcg}) and the abundance is determined by interpolating between the model points that brackets its location.  The models used for this work are those from \citet{mcg} in which the most massive star produced from the initial mass function has a mass of 60\,M$_{\odot}$. 

Each \hii\ region in our sample is plotted in Figure \ref{fig:mcg} and the McGaugh abundance models are overlaid on this plot.  These models are based on the behavior of the strong oxygen lines (i.e., the R$_{23}$ parameter) with changing metallicity which is well studied and has been calibrated to an oxygen abundance scale using both empirical and theoretical methods \citep[e.g.,][]{ep84, mcg, kd02, pil05, nag06}. The addition of the O$_{23}$ parameter allows for better determination of the ionization parameter which, in turn, allows a more accurate estimate of abundance \citep[e.g.,][]{kd02, nag06}. However, as can clearly be seen in Figure \ref{fig:mcg}, the relationship is not single-valued; there exists a ``high-metallicity" branch of the models where oxygen abundance decreases with an increase in R$_{23}$ and a ``low-metallicity" branch where oxygen abundance increases with an increase in the R$_{23}$ ratio.  In short, each point on the McGaugh models corresponds to two different possible metallicity determinations.

In order to break this degeneracy, the [\ion{N}{2}]/[\ion{O}{2}] ratio is used as a diagnostic to determine if the \hii\ region belongs on the ``high-" or ``low-" metallicity branch.  For this paper, the \hii\ region was placed on the low-metallicity branch if  [\ion{N}{2}]/[\ion{O}{2}]$< -1.0$ and was placed on the high-metallicity branch if [\ion{N}{2}]/[\ion{O}{2}] $> -0.9$. If $-1.0 <$ [\ion{N}{2}]/[\ion{O}{2}] $< -0.9$ the \hii\ regions was deemed to be in the ``turn-around region" where an average of the low- and high-metallicities was used.  Key line ratios and abundance results from the McGaugh method are presented in Table \ref{tab:mcg}.  The name of the \hii\ region is indicated in Column 1 and the calculated values of the strong line ratios R$_{23}$, O$_{23}$, and [\ion{N}{2}]/[\ion{O}{2}] are listed in 2 - 4.  Column 5 contains the oxygen abundance (log(O/H) $+$ 12) calculated for each \hii\ region using the McGaugh method, and Column 6 shows the difference between the McGaugh abundance and \te\ abundance where applicable.  

Notice that the difference between the two methods is not always trivial, averaging 0.08 dex with a maximum difference of 0.29 dex.   It has historically been known that these two methods can be discrepant beyond the assumed errors \citep[e.g.,][]{vz06, yin07, kew08, zah12}. There are additionally two factors that may account for relatively large differences between the abundances derived from the \te\ and McGaugh methods: (1) The McGaugh models used in this work are calibrated for an initial mass function where the most massive star produced in a star-forming episode is 60 M$_{\odot}$, which may not be accurate for all of the star-forming regions being considered. (2) Several of the \hii\ regions for which $\Delta$log(O/H) is large are those which lie near the turn-around region in the McGaugh models, thus making it inherently more difficult to determine an accurate abundance with the McGaugh method. 

Due to errors and intrinsic scatter, it is possible that the [\ion{N}{2}]/[\ion{O}{2}] for a given \hii\ region may place it on the incorrect branch of the McGaugh models; this problem is especially prominent near the turn-around region. In cases where we had a single galaxy with multiple \hii\ regions placed on different branches due to [\ion{N}{2}]/[\ion{O}{2}], we ``forced" all the \hii\ regions onto the branch that was deemed appropriate.  The appropriate branch was chosen by considering all of the \hii\ regions in the galaxy, with special attention to any \hii\ region with a \te\ abundance. In addition, the [\ion{N}{2}]/[\ion{O}{2}] values were examined to determine if they were near enough to the boundaries such that a moderate error could lead to the misclassification. This ``forcing" was not often necessary and each instance where this occurred is noted in Table \ref{tab:mcg}. 

Though we do not have explicit errors associated with the McGaugh abundances, we assume an error of 0.1 dex in log(O/H)+12.  This estimated error is roughly consistent with the scatter that has been observed via comparing strong-line abundances from the McGaugh method with direct abundance determinations in the literature and in this sample.  

\subsection{Abundance Results}
\label{subsec:abunres}
An average ``global" oxygen abundance for each galaxy was obtained by calculating the weighted average of the individual \hii\ region abundances in each galaxy.  The \te\ abundance was used for the 16 \hii\ regions listed in Table \ref{tab:elsa}, and the McGaugh abundances from Table \ref{tab:mcg} were used for the remaining 32 \hii\ regions.  The average derived abundance for the 10 galaxies considered here  (excluding the two outliers ADBS 113845+2008 and 125156+1205) was calculated to be (log(O/H)+12)$_{\textrm{av}}$ =  7.98. Table \ref{tab:av} lists the mean global oxygen abundance for each galaxy (where the error is given as the standard error on the mean) as well as the standard deviation and range of abundances of the individual \hii\ regions within the galaxy.

These abundance results are broadly consistent with the expectations for star-forming dwarf galaxies, except in the cases of ADBS 113845+2008 and ADBS 125156+1205 which both have oxygen abundances more than 0.7 dex higher than the average for the rest of the sample.  As previously mentioned, from the diagnostic diagram alone we can establish that ADBS 125156+1205 is most likely  an edge-on spiral, not a dIrr, which is consistent with its relatively high abundance.  ADBS 113845+2008 is  a more enigmatic case that is discussed in more detail in Section \ref{sec:bcd}. 

\section{ABUNDANCE ANALYSIS}
\label{sec:abun}

\subsection{Luminosity-Metallicity Relationship}
 We next investigate the nature of the luminosity-metallicity (L-Z) relationship for the galaxies in this paper and how it compares to similar trends for dIrrs observed by other authors \citep[i.e.,][]{Leq79, skill89, pil01,  lee03, lee032, mel04, trem04, vzh06}.  Our results are illustrated in Figure \ref{fig:metlum} where we have plotted M$_{\textrm{B}}$ vs. average global oxygen abundance (log(O/H)+12) for our sample and a sample taken from the literature. The galaxies from this study are plotted as filled black points, with the exception of ADBS 113845+2008 and 125156+1205 which are plotted as a filled triangle and \texttt{x}, respectively. The literature sample used for comparison was taken from \citet{vzh06} and \citet{lee06} and are plotted as open red triangles and open blue circles, respectively.  

We fit our data using a bivariate linear regression method that accounts for measurement errors in both x and y, following the method outlined in \citet{akber96} and references therein.  From this fit, we derive the following relationship between the luminosity and metallicity for the galaxies in this work:
\begin{equation}
\label{me}
\textrm{log(O/H)} + 12\,\,=\,\,(5.25\,\pm\,0.59) - (0.162\,\pm\,0.035)\,\textrm{M$_{\textrm{B}}$}.
\end{equation}
This relationship is graphically illustrated in Figure \ref{fig:metlum} with a bold black line. When calculating our fit we excluded the two obvious outliers (ADBS 113845+2008 and 125156+1205) for the reasons previously discussed.

In order to remove systematic distance differences due to differing distance scales between our sample and the literature sample we adjusted all velocity-based distances (and quantities calculated from distance) to be consistent with a with a Virgocentric infall model with H$_{\textrm{o}}$ = 70 km/s/Mpc (consistent with the velocity-dependent distances assumed for our sample). The \citet{lee06} sample all have velocity-independent distances, so no adjustment was needed. However, most of the distances in the \citet{vzh06} sample were determined using a flow model with H$_{\textrm{o}}$ = 75 km/s/Mpc, so we have adjusted those as needed.  Once properly adjusted, we fit the combined samples from \citet{vzh06} and \citet{lee06} using the same bivariate method used for fitting our data.  This resulted in the following relationship:
\begin{equation}
\label{lit}
\textrm{log(O/H)} + 12\,\,=\,\,(6.07\,\pm\,0.14) -  (0.120\,\pm\,0.009)\,\textrm{M$_{\textrm{B}}$},
\end{equation}
shown in Figure \ref{fig:metlum} as a thin dashed line with dotted lines above and below representing the 1$\sigma$ standard error in the intercept. This fit is roughly consistent with that published in \citet{lee06} but shallower than that found by \citet{vzh06}.

All but one of the galaxies from our sample fall \emph{below} the best fit line obtained from the combined \citet{vzh06} and \citet{lee06} samples, suggesting that the galaxies analyzed in this work are either ``under-abundant" or ``over-luminous" compared to the dIrrs in those studies.  While our slope is broadly consistent with that obtained from fitting the literature galaxies, it is slightly steeper, which may be a statistical effect caused by our small sample size and narrow range in M$_{\textrm{B}}$. Thus, in order to obtain a more consistent quantitative comparison between our sample and the literature sample, we additionally performed a bivariate weighted linear-least-squares fit to our data with the slope held constant at $-0.120$ (From Eq. (2)) and allowed only the intercept to vary. We derive an intercept of 5.95 via this method, which is 0.12 dex below the intercept obtained from the the literature data and just within the formal error on that intercept. The fit is shown in Figure \ref{fig:metlum} as a long-dashed line. It appears that, at a given luminosity, the galaxies in our sample have a lower in abundance than the comparison sample, however the intercept of both fits are consistent within the uncertainties.

We recognize the systematics that can be associated with using B-band luminosity as a stand in for stellar-mass when examining scaling relationships with metallicity.  These usually are attributed to differing mass-to-light ratios, uncertainties due to internal absorption, and B-band enhancements due to starbursts.  

Since, the metallicity-luminosity relationship is often believed to more fundamentally be a relationship between mass and metallicity, it is common to use an estimate of the stellar mass to avoid some of these systematics. We used the method of \citet{bd00} to calculate an estimate for the the stellar mass using the B band luminosity and B-  V color.  However, when we examined the resulting mass-metallicity relationship we found that it exhibits a larger degree of scatter than the luminosity-metallicity relationship. This increased scatter is most likely because our photometric data consists of only broadband B and V observations, thus making accurate mass estimates difficult to calculate precisely.  For this reason, we are choosing to limit our analysis to the L-Z relationship, with the realization that there is likely some scatter due to differing mass to light ratios.

We should not be strongly effected by B-band luminosity enhancements from starburst as all of the galaxies we consider are relatively low-surface brightness (with the exception ADBS 113845+2008) and thus the enhancements due to starburst episodes should be small \citep{lee04}. Additionally, all of the galaxies used here are actively star forming galaxies with emission-line nebulae and similar morphology (dIrrs), thus there should be no systematic luminosity offsets between the samples.  We expect very little uncertainty associated with dust absorption since each of our galaxies has a relatively low c(H$\beta$) (with the exception of ADBS 125156+1205, the edge-on spiral) suggesting very little dust absorption is present.

The two outliers that were excluded from the majority of our analysis (ADBS 125156+1205 and 113845+2008) both lie well \textit{above} the best fit line in all cases, and therefore appear to be significantly over-abundant or under-luminous in comparison. As mentioned previously, ADBS 125156+1205 is likely an edge-on spiral for which the luminosity has been under-estimated since we have not accounted for internal absorption and have no robust way to do so. Therefore, the luminosity measurement presented here should be considered a lower-limit for this galaxy. The position of ADBS 113845+2008 on Figure \ref{fig:metlum} is more difficult to understand.  If the observed offset of ADBS 113845+2008 is real, we must assume that it is due to a relative ``over-abundance."  In order to account for the offset in luminosity, the M$_{\textrm{B}}$ measurement would need to be too dim by more than 4 magnitudes. Even if we assume there is a very extended, extremely low-surface brightness component to this galaxy (which has not yet been detected), an increase of 4 in the measured magnitude is unreasonable, so we can conclude that the offset is much more likely due to an abundance anomaly.  Despite being omitted as an ``outlier" for much of this paper, ADBS 113845+2008 is an interesting galaxy and its anomalous properties may be important in achieving a deeper understanding of the chemical evolution of galaxies embedded in extended gas disks. We devote Section \ref{sec:bcd} to discussing the nature of this galaxy.

\subsection{Relative Abundances of N and $\alpha$-process Elements} 
\label{sec:rel}
Oxygen abundance is used as an indicator of metallicity because reasonable estimates of the oxygen abundance can be made even in cases where only the strong lines are measurable.   If the temperature sensitive \oiiite\ line was detected, we calculated and tabulated the relative abundances of other $\alpha$-process elements (Ne, S, Ar) and nitrogen  (Table \ref{tab:elsa}) in addition to the oxygen abundance .  Since the $\alpha$-elements are all synthesized under similar conditions in massive stars, the relative abundance of these elements is expected to remain constant, independent of the metallicity of the \hii\ region as empirically established by many authors \citep[i.e][]{vz97, it99, mel04, vzh06}.  In other words, we expect the ratios of Ne/O, S/O, and Ar/O to remain roughly constant with increasing oxygen abundance.   

Figure \ref{fig:other}\emph{(a)-(c)} shows the relative abundances of Ne, S, and Ar vs. metallicity for the \hii\ regions in Table \ref{tab:elsa} (black points).  These plots additionally contain a point representing ADBS 113845+2008\emph{a} (blue triangle) which we considered separately below. 

Each section of Figure \ref{fig:other} contains a dotted-horizontal line representing the average abundance ratio of this sample.  Our results show that the $\alpha$-element abundance ratio remains roughly constant with increasing metallicity as expected. However, the average Ar/O and S/O values for this sample are slightly lower than those found in the literature.  Averages from this study, as well as others selected from the literature \citep{it99, mel04, vzh06} are presented in Table \ref{tab:other}. 

Figure \ref{fig:other}\emph{(d)} shows the abundance of nitrogen relative to oxygen for the \hii\ regions in Table \ref{tab:elsa}.  Nitrogen is a somewhat complex case because the CNO cycle in intermediate- and low-mass Population I stars can additionally contribute to the enrichment of N.  In low-metallicity systems it is expected that the nitrogen to oxygen ratio should remain roughly constant with increasing oxygen abundance since nitrogen is expected to be a primary element synthesized from the pristine H in these low-metallicity nebulae \citep[e.g.][]{vz98, it99, mel04}.  But, as illustrated in \citet{vz98}, as log(O/H)$+$12 increases significantly above $\approx8.3$, the contribution of N  generated from the CNO cycle in lower-mass stars becomes important, leading to increasingly N-enriched star-forming nebulae in high-metallicity systems.  

As can be seen in Figure \ref{fig:other}\emph{(d)}, the N/O ratio appears to remain constant across the range of oxygen abundances for most of our sample, which is consistent with our expectations given that all the \hii\ regions included in Table \ref{tab:elsa} are low-metallicity systems.  What is of more interest in Figure \ref{fig:other}\emph{d} is the location of the blue triangle that represents ADBS 113845+2008\emph{(a)}.

The abundance ratios for ADBS 113845+2008\emph{a} were determined similar to the \hii\ regions in Table \ref{tab:elsa}. However, since the spectrum of ADBS 113845+2008\emph{a} does not contain a measurable \oiiite\ line, the electron temperature was estimated using the oxygen abundance determined with the McGaugh method.  This resulted in an assumed \te $\approx$ 7,000 K. The abundances obtained for ADBS 113845+2008\emph{a} via this deductive approach are listed in Table \ref{tab:other}. 

It appears that nitrogen is significantly enhanced in ADBS 113845+2008\emph{a} compared to the average of the low-metallicity \hii\ regions (Figure \ref{fig:other}\emph{d}). If ADBS 113845+2008 is truly a metal-rich galaxy, as indicated by the derived McGaugh abundance, than an enhanced nitrogen abundance is expected, following the results of \citet{vz98}. We note that it appears ADBS 113845+2008 is slightly over-enriched in each $\alpha$ element included on Figure \ref{fig:other}, which could be because \te\ was underestimated resulting in a higher assumed metallicity, but also could be due to errors in the ionization correction factor associated with the relatively uncertain temperature.  Even though all the $\alpha$ elements appear slightly over-enriched, in all cases other than nitrogen, the relative abundances are consistent with the average within the errors. We chose to include ADBS 113845+2008\emph{a} on Figure \ref{fig:other} primarily to explore whether the derived N abundance was consistent with the galaxy being relatively metal-rich (as indicated with the McGaugh analysis) and caution that the \te\ abundance results presented for for this galaxy be taken only as estimates.

\section{DISCUSSION}
\label{sec:disc}

\subsection{Metallicity Luminosity Trend}
The relationship between luminosity and metallicity for the dIrrs in this sample is consistent with the previously established result that metallicity (given as log(O/H)) increases more or less linearly with increasing luminosity.  A bivariate least squares fit to our sample (Eq. 1) produces a slope that is broadly consistent with the literature but somewhat steeper than a similar fit to the larger literature-based sample composed of dIrrs from \citet{vzh06} and \citet{lee06} that is given in Eq. 2.  We recognize that the steepness of our slope is likely related to our relatively small sample size and the small range of M$_{\textrm{B}}$ covered by our data.  When the slope from the literature sample is assumed and our sample is refitted, we achieve a similar reduced $\chi^{2}$ and still find compelling evidence that our sample deviates from the L-Z relationship derived from the literature sample. Compared to the galaxies from the literature, our galaxies are generally under-abundant for their derived luminosity.

Our sample was selected in a different manner than samples which have been previously used to establish the L-Z relationship, thus it is plausible that the relative abundance offset we see is due to an underlying physical difference between our sample and those more often analyzed.   In this section we investigate several possible ``second parameters" that may be responsible for some of the intrinsic scatter in the metallicity-luminosity relationship, and also may help us understand the differences between our sample and those taken from \citet{vzh06} and \citet{lee06}.

\subsection{Second Parameter Analysis}

In Figure \ref{fig:res_fig} we have plotted the residuals in oxygen abundance vs. four potential second parameter quantities:  (a) dynamical mass (M$_{\textrm{dyn}}$), (b) gas richness (M$_{\textrm{HI}}$/L$_{\textrm{B}}$), (c) current star formation rate (SFR) and (d) central-surface-brightness ($\mu_{B,0}$). To calculate the residual in log(O/H) we used Equation 2 (dashed line in Fig. \ref{fig:metlum}) to determine the ``fiducial" metallically.

\subsubsection{Dynamical Mass}

Figure \ref{fig:res_fig}(a) indicates a suggestive negative-correlation between the oxygen-abundance residual and M$_{\textrm{dyn}}$ (estimated from the global 21cm line width; see Table \ref{tab:basic}). When these data were fit with a weighted linear-least-squares regression, the relationship was found to be statistically significant with a slope of -0.22$\,\pm\,$0.10.  This suggests that the most massive galaxies from our sample are also those that are the most under-abundant at a given luminosity.  This is contrary to the simple ``open-box" model of galaxy evolution that is often invoked to explain the L-Z relationship.  In the open-box model, the L-Z relationship is better interpreted as a metallicity-mass relationship which develops because massive galaxies are more efficient at retaining enriched supernova ejecta. To be consistent with this model, one would expect the more massive galaxies to be over-enriched from the fiducial abundance at a given luminosity. Our finding is consistent with a similar suggestive trend between abundance residual and M$_{\textrm{dyn}}$ mentioned in \citet{vzh06}.

\subsubsection{Gas-Richness and Star-Formation}

Modestly significant negative-correlations also appear in Figures \ref{fig:res_fig}(b) and (c), indicating the most under-abundant galaxies correlate with high-values of gas richness and current star formation rate as well.  When fit with a weighted linear-least-squares regression, the trend with gas-richness was found to be statistically significant, similar to the trend in Figure \ref{fig:res_fig}(a), with a slope of -0.28$\,\pm\,$0.11 while the trend with star-formation rate was only marginally significant (less than 2$\sigma$) with a  slope of -0.10$\,\pm\,$0.06.  

\emph{Gas-richness}\\
The trend with gas-richness is perhaps one of the most significant findings.  Since these galaxies were chosen from a 21cm catalog, we have well determined gas masses for each galaxy and are able to analyze our results in the context of their gas content.  This is particularly important for understanding the evolution of these galaxies and the importance of factors such as gas infall and outflow. The trend observed with gas-richness may indicate that pristine gas has diluted the enriched gas in these galaxies due to infall or mixing of \hi\ from an extended disk.  This conclusion is consistent with the negative correlation with dynamical mass observed in Figure \ref{fig:res_fig}(a) as well.  The mechanism that would trigger infall or mixing events in a relatively isolated dwarf galaxy is unclear, however one would expect that galaxies with large dynamical masses, and thus deeper potential wells, would be more likely to experience such infall events.

In the context of the ``passive evolution" scenario outline by \citet{greb03}, where it is suggested that gas-rich dIrrs are the less evolved progenitors of gas-poor dwarf spheroildals, the trend with gas-richness could possibly indicate that these galaxies are ``less evolved." However, we find this interpretation unconvincing for our data.  The galaxies we examined have moderate star-formation rates (log(SFR) = -1.69 -- -0.67 M$_{\odot}$ yr$^{-1}$ ) and relatively large \hi-masses (log(M$_{\textrm{HI}}$ = 8.21 -- 9.67 M$_{\odot}$), thus these galaxies could not feasible exhaust their gas supply in a Hubble Time without a dramatic spike in the SFR or significant stripping of the gas content.  It is highly unlikely that these gas-rich dIrrs are likely to evolve into gas-poor dwarf spheroidals on a timescale that is shorter than the Hubble Time.  

Compared to the literature samples, our sample is only moderately gas rich despite being chosen from an \hi\ survey. Our galaxies have a median M$_{\textrm{HI}}$/L$_{\textrm{B}}$ of 1.16 M$_{\odot}$/L$_{\textrm{B},\odot}$ which is significantly more gas rich than the \citet{lee06} sample (median M$_{\textrm{HI}}$/L$_{\textrm{B}}$ = 0.01M$_{\odot}$/L$_{\textrm{B},\odot}$) , but less gas rich than the sample from \citet{vzh06} (median M$_{\textrm{HI}}$/L$_{\textrm{B}}$ = 1.40 M$_{\odot}$/L$_{\textrm{B},\odot}$). The fact that all but one of our galaxies appear to be ``under-abundant" despite being less gas rich on average than the comparison sample indicates that gas richness is not the predominate second parameter causing the differentiation between our sample and that from the literature. 

If we additionally consider the total \hi-mass and look at M$_{\textrm{HI}}$ for our sample versus the literature samples, we see that the median M$_{\textrm{HI}}$ for our sample is much larger than that from either \citet{vzh06} or \citet{lee06}.  Our data are limited by \hi\ flux and since our average distance is much larger it is a natural extension that our average \hi\ mass is larger as well. The flux limit of the ADBS survey reached down to approximately 10$^{8}$ M$_{\odot}$, in most cases, and our sample was chosen from those with the lowest \hi\ masses producing a median \hi\ mass of log(M$_{HI}$) = 8.97 M$_{\odot}$, while the literature samples have median \hi\ masses of only log(M$_{HI}$) = 8.39 M$_{\odot}$ \citep{vzh06} and log(M$_{HI}$) = 8.09 \citep{lee06}.  

Perhaps the fact that our galaxies are both gas rich \emph{and} contain a very large \hi-reservoir is important to consider. It is possible that many of theses galaxies may be imbedded in larger \hi-disks, and it could be the presence of those disks that has influenced the chemical evolution of our galaxies (through processes such as the mixing of pristine \hi) compared to a sample from the literature that was chosen in a different manner.  When considering M$_{\textrm{HI}}$ as a second parameter we compared the oxygen-abundance residual vs. \hi-mass and found a similarly significant negative correlation to that observed with M$_{\textrm{dyn}}$ but with a shallower slope (-0.11$\pm$ 0.05).

\emph{Star-Formation Rate}\\

The slight negative-correlation with star-formation rate indicates that the most under-abundant galaxies are currently producing stars very efficiently. This is consistent with suggestion that the infall of pristine gas may be an important factor in the evolution of the gas in these system. If gas infall occurs, it may trigger star formation events by increasing the surface density of \hi\ above the critical density for stability \citep{ken98}.  It is possible that the marginal trend with SFR appears in part from slight B-band luminosity enhancements in the galaxies with the highest rates of star-formation \citep{lee04, salz05}. We return, however, to the argument that our sample is composed of mostly low-surface brightness galaxies, where the 3 most under-abundant are all low-surface brightness ($\mu_{\textrm{0,B}} >$ 21.75; Fig \ref{fig:res_fig}(d)), therefore we do not expect to see notable luminosity enhancements due to star-formation in these systems. Even if we assume that the slight anti-correlative trend with SFR is a result of luminosity enhancements, the lack of a positive correlation still suggests that the abundance offsets from the literature sample are not the result of inefficient star formation.

\subsubsection{Central Surface Brightness}
 
Figure \ref{fig:res_fig}(d) displays no significant correlation between abundance residuals and central-surface-brightness.  This was confirmed with a weighted linear-least-squares fit that returned a slope consistent with zero within the errors. The lack of trend is notable because it has been suggested that low surface brightness galaxies may be less evolved than their higher surface brightness counterparts \citep{mcg94} which would lead to a positive correlation.  While we do not see any statistically significant trend in Figure \ref{fig:res_fig}(d), we do note that three of the four most under-abundant galaxies have low central-surface-brightness values, and thus do not completely exclude the possibility that a larger sample of galaxies may display a notable trend.

The median central-surface-brightness value for our sample ($\mu_{0,B}$ = 21.8) is brighter than the median of the \citet{vzh06} sample ($\mu_{0,B}$ = 22.7), indicating that, while our sample is relatively low surface brightness, on average the \citet{vzh06} sample consists of lower surface brightness galaxies.  The surface brightness for the \citet{lee06} was not measured and cannot be compared.  However, our analysis indicates that is is unlikely that surface brightness is a distinguishing second parameter between our sample and that taken from the literature.

\subsubsection{Other Parameters}

In addition to the four second parameters already discussed, we examined several other possible second parameters not shown in Figure \ref{fig:res_fig}.  We looked at the oxygen abundance residuals vs. M$_{\textrm{HI}}$, color (B$-$V),  and surface density of \hi\ and found no significant correlations with color or surface density of \hi. As stated previously, a trend similar to that observed with M$_{\textrm{dyn}}$ but with a shallower slope (-0.11$\,\pm\,$0.05) was found when the oxygen-abundance residual was compared with \hi\ mass.  We also investigated whether alternative metrics for gas richness or SFR provided further insight. When ``gas richness" is estimated as M$_{\textrm{HI}}$/M$_{\textrm{dyn}}$ no statistically significant trend was found, and when the gas richness parameter was instead calculated as M$_{\textrm{HI}}$/M$_{*}$ (with M$_{*}$ calculated following \citet{bd00}) we find a negative correlation with similar significance to that found with M$_{\textrm{HI}}$/L$_{\textrm{B}}$ but with a shallower slope (-0.12$\,\pm\,$0.05).  Finally, when we normalize the SFR by the M$_{*}$ to analyze the oxygen abundance residual vs. specific star formation rate, we find a trend that is nearly identical to that seen with the non-normalized SFR in Figure \ref{fig:res_fig}(c). 

\subsubsection{Implications from Second Parameter Analysis}

While we are not able to make any conclusive statements concerning the role of any one observational second parameter in the metallicity luminosity relationship, the observed negative-correlations with gas richness, dynamical mass, and star-formation rate together indicate that infall and/or mixing of pristine gas may play a significant role in the evolution of these galaxies.  The relationships observed with dynamical mass and star-formation rate support the interpretation that the infall or mixing of pristine gas may be leading to dilution and therefore lower abundance.  A large dynamical mass, and thus deeper potential well, makes an infall event more likely to occur. Additionally, if these galaxies have experienced recent infall or mixing of \hi, the enhanced star-formation rates that we observe would also be expected. It is difficult to draw broad conclusions from this small sample, but we see indications of gas infall and mixing affecting the evolution of gas-rich dIrr galaxies

While our sample, on average, is no more gas-rich or low surface brightness than the \citet{vzh06} sample, we do note that the three \textit{most} under-abundant galaxies, ADBS 115906+2428, 145647+0930, and 153703+2009, are the three most gas-rich galaxies in our sample ($M_{HI}$/$L_{B}$ $>$ 1.51) and additionally have low central surface brightness ($\mu_{0,B} > $ 21.5) as well as relatively large dynamical mass (log(M$_{\textrm{dyn}}$) $>$ 10.3; see Fig. \ref{fig:res_fig}). UGC 10445, the most ``under-abundant" galaxy from the \citet{vzh06} sample (i.e. the galaxy with the most negative abundance residual) shares these same characteristics of having a large dynamical mass (log(M$_{\textrm{dyn}}$) = 10.24; the largest in that sample), low surface brightness ($\mu_{0,B}$ = 21.8), and is relatively gas-rich (log($M_{HI}$/$L_{B}$) = 0.0).

\subsection{Metallicity Variation in Dwarf Galaxies}

Since we obtained spectra for multiple \hii\ regions in most of our galaxies, we are able to examine our results for indications of metallicity variation within these galaxies.  In Figure \ref{fig:ind} we have plotted the individual \hii\ regions from four of our galaxies (differentiated by color and symbol) onto a diagnostic diagram. These four galaxies were chosen because they each contain several \hii\ regions with measured oxygen abundances and at least one \hii\ region with a direct abundance determined using the \te\ method. 

Notice that the \hii\ regions from ADBS 125850+1308, 135822+2533, and 145647+0930 are spread over a fairly large region of the diagram, including a significant spread along model star-forming nebulae line.  The metallicity from the models of \citet{de86} increases smoothly along this line, with the most metal-poor star-forming regions at the top left, and the most metal-rich at the bottom right.  Thus, the observed spread of our points would seem to indicate a significant metallicity variation within these galaxies.  However, the results from our abundance analysis indicate that the metallicity is relatively uniform within each galaxy, consistent with the well-established result that dwarf galaxies do not display significant metallicity gradients \citep[e.g.,][]{ks96, ks97, lsv06, crox09}.

There are slight abundance variations between \hii\ regions within several of our galaxies, but these variations are, in general, not large enough to be significant, considering the intrinsic error associated with the measurements.  The one exception is ADBS 125850+1308 where the abundance measurements range over 0.27 dex.  The significance of this variation is not totally clear, because the standard deviation about the mean is only 0.103 dex which is consistent with the assumed error on the McGaugh abundances. All but a few \hii\ regions are consistent, within the cited errors, with the global average (log(O/H)+12 = 8.05) from Table \ref{tab:av}.  The exceptions are ADBS 125850+1308\emph{c, f, h, \textrm{and} i} which represent the two lowest (\emph{c, h}) and highest (\emph{f, i}) abundance \hii\ regions in the galaxy.  It is interesting that ADBS 125850+1308\emph{c} and \emph{i} both have \te\ abundances, thus smaller errors, yet represent the two \hii\ regions with the largest deviations from the mean. It seems plausible that there is some metallicity variation throughout ADBS 125850+1308, however it is difficult to say with certainty given the relatively large uncertainty associated with the McGaugh abundances used for many of the \hii\ regions. 

If we examine the finder chart for this galaxy in Figure \ref{fig:finder}, there is no obvious systematic nature to this variation. However, ADBS 125850+1308\emph{i}, the most metal-poor \hii\ region, does appear considerably isolated from the rest of the star formation activity in the galaxy, consistent with its lower abundance. There are several other notable properties of ADBS 125850+1308 that may also relate to the observed metallicity variations. ADBS 125850+1308 is \textit{very} red for a dIrr (B-V = 0.65), as well as fairly gas poor (log(M$_{\textrm{HI}}$/L$_{\textrm{B}}$) = -0.120).  The color composite image from the Sloan Digitized Sky Survey shows a very blue section in the southeast region of the galaxy while the majority of the system appears significantly more red.  This indicates the system may be a recent merger or perhaps just a chance projection of a very blue dwarf onto a larger, redder galaxy. The bright \hii-region (region \emph{i}) in the northwest portion also appears very blue, yet does not appear connected to the larger blue portion. Some of the low-metallicity \hii\ regions in this study do appear associated with the bluer section, but, due to the way which are slit positions cut through the galaxy, it is not totally clear that the blue region has a distinctly different abundance than the redder portion.  We note that all the \hii\ regions appear are at very nearly the same velocity, consistent with this being a single object.

The standard deviation for the \hii\ regions in ADBS 135822+2533 is very small (0.051 dex) and the range is similarly small at only 0.10 dex.  Both of these are less than the assumed error on the majority of the abundance measurements indicating the oxygen abundance is quite uniform throughout this galaxy.  \hii\ regions from ADBS 145647+0930 cover a larger range of 0.22 dex, but the standard deviation (0.071 dex) is still significantly less than the typical error in a single measurement, indicating the oxygen abundance is relatively uniform in this galaxy as well.  The \hii\ regions in ADBS 153703+2009 additionally span a narrow range (0.16 dex) and have a standard deviation (0.066 dex) well within the assumed errors, indicating uniform metallicity throughout the galaxy.  Other galaxies that contain multiple \hii\ regions also have quite uniform oxygen abundances as indicated in Table \ref{tab:av}.  ADBS 115906+2428 contains only two \hii\ regions, but they have significantly different oxygen abundances (spanning 0.21 dex; both from \te\ measurements).   While this is intriguing it is difficult to quantify the significance without another \hii\ region for comparison.

\subsection{ADBS 113845+2008}
\label{sec:bcd}
We have re-analyzed the spectrum for ADBS 113845+2008\emph{a}, which was included in \citet{can1138} as part of a study that performed an in-depth analysis of the neutral gas structure and dynamics of ADBS 113845+2008.  Using the McGaugh method, we have calculated an updated oxygen abundance that indicates that ADBS 113845+2008\emph{a} may be significantly over-enriched for its luminosity.  Using the luminosity-metallicity trend from our literature sample, we would expect a log(O/H)+12 = 8.05, while we derive an abundance using the McGaugh method of log(O/H)+12 = 8.71 from our spectral data.  

Similar outliers to the L-Z relationship have been found by other authors.  Notably, \citet{peep08} and \citet{zah12} investigate a set of low-luminosity high-metallicity outliers found in the Sloan Digital Sky Survey.  \citet{berg11} suggest that many of the outliers discussed in the \citet{peep08} may not truly be outliers, but instead are galaxies whose oxygen abundance was overestimated due to biased strong-line methods, yet \citet{zah12} contend that while some galaxies may be affected by such a bias, there remains a significant set of true outliers.  We do not have any reason to believe that the oxygen abundance we determined for ADBS 113845+2008 suffers any such bias, but since there is no direct measurement of the \oiiite\ line to compare to, it is difficult to say with certainty.  

The outlier galaxies analyzed in \citet{peep08} have morphologies that are not dissimilar from ADBS 113845+2008.  The optical broadband images of the \citet{peep08} sample suggest that the outlier galaxies have dSph or dwarf elliptical (dE) like morphologies that are undisturbed by any recent interaction.  Additionally, 10 of the 41 galaxies analyzed in that work have very bright or very blue cores.  ADBS 113845+2008 optically has a very symmetric morphology that shows no evidence of recent interaction and has an extremely compact bright blue core.  We would hesitate to classify ADBS 113845+2008 as a dSph or dE due to the compact nature, but note that the morphology ADBS 113845+2008 is broadly consistent with those seen in the \citet{peep08} outliers.  

\citet{peep08} suggest that their outliers are likely gas-poor transition galaxies, which is certainly not true for ADBS 113845+2008.  ADBS 113845+2008 has a relatively large gas fraction (log(M$_{\textrm{HI}}$/L$_{\textrm{B}}$ = 0.048 M$_{\odot}$/L$_{\odot}$) and is imbedded in an enormous \hi\ disk \citep{can1138}.  Since the work of \citet{peep08} did not contain a measurement of the gas content for their galaxies, it would be interesting to investigate if they are truly gas-poor as suggested in that paper, or if they have gas content and gas morphologies more similar to ADBS 113845+2008.  

It was the conclusion of \citet{can1138} that this galaxy is a rare object within the local universe which has evolved in relative isolation, maintaining ongoing star-formation only in a very compact inner region.  They find that this star-formation is most likely the result of stochastic processes, such as turbulence, in the disk, but only the inner-most regions have achieved sufficient densities to initiate star-formation. Our result, that ADBS 113845+2008 appears to be a metal-enriched galaxy with a much higher abundance than expected for its luminosity supports their conclusion that it is not a young object.  In fact, judging by the oxygen abundance obtained from the strong line method it appears to have been relatively efficient at forming stars over a Hubble-time, but only within a limited portion of the gas disk. 

If ADBS 113845+2008 is truly metal rich, how it has managed to maintain its enriched gas so well is not clear.  \citet{can1138} derived a relatively large dynamical mass of log(M$_{\textrm{dyn}}$) = 10.58 for this galaxy, however, this is not beyond the range of other dwarf-irregulars included in this study, and does not explain how ADBS 113845+2008 could be so dramatically more effective at maintaining enriched gas compared to other galaxies with similar dynamical masses.  It seems more likely that its ability to maintain enriched gas must be related to the highly centrally-concentrated nature of the star formation in this galaxy. 

The highly centralized nature of the star-formation activity may also explain why ADBS 113845+2008 has experienced no significant mixing of the pristine gas, while, as hypothesized in Section 7.1, similar systems may show signs of infall and mixing.  Feedback from star-formation and supernova occurring throughout a significant portion of the disk may be necessary to enable the mixing of pristine gas, or, alternatively, the instabilities that initialize star formation throughout the disk could be necessary to initialize mixing as well.  The stability of the gas disk and lack of even low-level star-formation throughout the disk of ADBS 113845+2008 may mean that the conditions for mixing were never met so any infalling pristine gas has remained unmixed, or, perhaps ADBS 113845+2008 has never experienced any significant infall events.  It seems likely that the unusual structure of this galaxy is linked to its unusual chemical enrichment history, but the specifics remain a mystery, and we echo the conclusion of \citet{can1138} that ADBS 113845+2008 appears to be a truly unique object in the local universe.

\section{CONCLUSIONS}
\label{sec:con}

We have determined the nebular abundances for 48 \hii\ regions in 12 low-mass galaxies selected from the Arecibo Dual Beam Survey \citep{rs2000}.  Our sample is composed primarily of star-forming dIrrs but also included an edge-on spiral (ADBS 125156+1205) and a compact dwarf galaxy (ADBS 113845+2008).   We obtain reliable measurements of the temperature sensitive \oiiite\ line for 16 \hii\ regions and thus are able to directly determine the electron temperature for these nebulae and use this to calculate the nebular abundances.  For the remaining 32 \hii\ regions we used the strong-line method from \citet{mcg} to determine the oxygen abundance. 

As part of our analysis we determined chemical abundance for the compact dwarf galaxy, ADBS 113845+2008, adding to the unique nature of this galaxy. The nebular abundance of the star-forming region in ADBS 113845+2008 determined via the strong-line method from \citet{mcg} (log(O/H) +12 = 8.71) is very high for its luminosity but could be consistent with the interpretation of \citet{can1138} that star-formation has occurred consistently in the innermost regions of ADBS 113845+2008 for nearly a Hubble Time. This result however raises further questions about the dynamical evolution of ADBS 113845+2008, including how it has so efficiently retained enriched gas while other similar luminosity systems are so metal poor.

The 10 ``normal" star-forming dIrr galaxies in our sample were determined to have an average oxygen abundance of log(O/H)+12 = 7.98. These galaxies appear to follow a luminosity-metallicity trend that is broadly consistent with that established by previous authors.  When compared with samples of star-forming dIrrs taken from the literature \citep{lee06, vzh06}, however, our sample appears to be under-abundant at a given luminosity. 

We analyzed several possible ``second parameters" to determine if the cause of the relative under-abundance could be identified.  This analysis identified suggestive correlations between relative under-abundance and dynamical mass, \hi\ mass, gas richness, and star-formation rate. Additionally, while we found no statistically significant correlation with central-surface-brightness, the three galaxies that appear the most under-abundant for their luminosity all have relatively low central surface brightness values. Our analysis provided no definitive second parameter, but instead has indicated that multiple second parameters appear to be  important.  Specifically, the observed correlations with gas richness, large dynamical mass, and elevated star formation rate have led us to suggest the infall and/or mixing of pristine \hi\ gas as a plausible cause of the apparent under-abundance observed for many of our galaxies.  

\acknowledgements
We would like to thank the anonymous referee for the constructive comments as well as the staff of the MMT Observatory for their excellent support during our observing runs.  JJS and NCH were supported in part NSF Grant AST-0708230.  In addition, NCH received financial support from an Indiana University Dissertation Year Fellowship and a Graduate Fellowship from the Indiana Space Grant Consortium.  

\clearpage



\clearpage
\scriptsize
\centering
\begin{landscape}
\begin{deluxetable*}{c c c c c c c c c c c c c c}
\caption{Galaxies Included in Sample} \label{tab:basic} \\
\hline \hline \noalign{\smallskip}
Galaxy & RA & Dec & Optical Catalog & m$_{\textrm{B,obs}}$ & A$_{\textrm{B}}$ & Distance & M$_{\textrm{B}}$ & B$-$V & $\mu_{0,\textrm{B}}$ & log(M$_{\textrm{Dyn}})$ & log(M$_{\textrm{HI}}$) & log(M$_{\textrm{HI}}$/L$_{\textrm{B}}$) &  {log(SFR)} \\
ADBS & (J2000) & (J2000) & & [mag] & [mag] & [Mpc] & [mag] & [mag] & [mag arcsec$^{-2}$] & [M$_{\odot}$] & [M$_{\odot}$] & [M$_{\odot}$/L$_{\odot}$] & [M$_{\odot}$ yr$^{-1}$]  \\
(1) & (2) & (3) & (4) & (5) & (6) & (7) & (8) & (9) & (10) & (11) & (12)  & (13) & (14)\\
\noalign{\smallskip} \hline \noalign{\smallskip}
071352+1031 & 07:13:52.1 & 10:31:26.4 & UGC3755          & 14.10$^{a}\,\pm\,$0.20 &  0.384 & 7.4$^{b}$ & -15.63$\,\pm\,$0.13 & 0.55$^{c}$ &   21.64$^{c}$ & 10.2 & 8.21 & -0.229 & -1.69 \\	 
081821+2431 & 08:18:21.1 & 24:31:37.2 & IC2271	         	    & 15.77$\,\pm\,$0.03     &  0.144 & 33.4      & -16.99$\,\pm\,$0.11 & 0.34  &   18.41  	    & 10.1 & 9.09 &  0.109 & -1.21 \\	 
100508+2207 & 10:05:08.9 & 22:07:37.2 & UGC5433          & 16.93$\,\pm\,$0.04     &  0.176 & 59.8      & -17.13$\,\pm\,$0.12 & 0.44  &   21.64        & 10.4 & 9.23 &  0.188 & -1.12 \\	 
100735+1306 & 10:07:35.3 & 13:06:36.0 & LSBCD637-18  & 16.89$\,\pm\,$0.17     &  0.163 & 41.9      & -16.39$\,\pm\,$0.12 & 0.34  &   22.84        & 9.25 & 8.76 &  0.014 & -1.61 \\
\end{deluxetable*}
\normalsize
\clearpage
\end{landscape}

\clearpage
\scriptsize
\centering
\begin{deluxetable*}{c c c c c c c }
\caption{Background Galaxies and Quasars} \label{tab:hiz} \\
\hline \hline \noalign{\smallskip}
Associated ADBS Image & Name & Type & RA & Dec & Cross ID & z \\
(1) & (2) & (3) & (4) & (5) & (6) & (7) \\
\noalign{\smallskip} \hline \noalign{\smallskip}
100508+2207 & B1 & QSO & 10:05:08.9 & 22:07:07.6 & SDSS J100508.91+220707.6  &  1.588$\,\pm\,$0.005 \\	 
100735+1306 & B2 & Galaxy & 10:07:33.2 & 13:06:38.7 & GMBCG J151.88842+13.11.076$^{e}$  &  0.388$\,\pm\,$0.0001\\	 
153703+2009 & B3 & Galaxy & 15:37:10.5 & 20:09:01.5 & SDSS J153710.46+200901.4 & 0.143$\,\pm\,$0.0001\\	 
\noalign{\smallskip} \hline
\hline 
\\
\multicolumn{6}{l}{$^{e}$\citet{hao10}}\\
\endfoot  
\normalsize
\clearpage
\end{deluxetable*}


\clearpage
\scriptsize
\begin{landscape}
\begin{longtable}{l c c c c c c c }
\caption{Line Measurement Results} \label{tab:gold} \\

\hline \hline \noalign{\smallskip}
Line Identification & 071352+1031\,\emph{a} &  071352+1031\,\emph{b} & 081821+2431\,\emph{a} & 081821+2431\,\emph{b} & 100508+2207\,\emph{a} &100735+2306\,\emph{a} & 113845+2008\,\emph{a} \\
(\AA) & {F$_{\textrm{ion}}$/H$\beta$} & {F$_{\textrm{ion}}$/H$\beta$} & {F$_{\textrm{ion}}$/H$\beta$} & {F$_{\textrm{ion}}$/H$\beta$} & {F$_{\textrm{ion}}$/H$\beta$} & {F$_{\textrm{ion}}$/H$\beta$} & {F$_{\textrm{ion}}$/H$\beta$} \\
\noalign{\smallskip} \hline \noalign{\smallskip}
$[$\ion{O}{2}$]\,\lambda 3728$   	      &	2.152$\,\pm\,$0.101  & 1.673$\,\pm\,$0.074   & 3.386$\,\pm\,$0.154  &	3.939$\,\pm\,$0.184  & 2.198$\,\pm\,$0.073   & 3.099$\,\pm\,$0.162 & 2.877$\,\pm\,$0.129  \\
H$\eta$    	                                  		      &	      --	   			 & 0.022$\,\pm\,$0.002   &	   --	      			 &		 --        		 &        --           		      &       -- 	     &      --            \\  
$[$\ion{Ne}{3}$]\,\lambda 3869$   	      &	0.369$\,\pm\,$0.018  & 0.354$\,\pm\,$0.014   & 0.295$\,\pm\,$0.014  &	0.196$\,\pm\,$0.012  & 0.343$\,\pm\,$0.019   & 0.245$\,\pm\,$0.017 & 0.145$\,\pm\,$0.007  \\
\ion{He}{1}+H$\zeta$  			      &       --	   			 & 0.129$\,\pm\,$0.005   &         --	      		  & 	 --        			 &        --           		      &       --  	     &      --            \\
$[$\ion{Ne}{3}$]\,\lambda 3970$+H$\epsilon$          &	      --	 & 0.176$\,\pm\,$0.007   &	   --	      			  &		 --        		 &        --           		      &       --  	     &      --            \\
\end{longtable}
\end{landscape} 


\scriptsize
\begin{deluxetable*}{l c c c c c c c }

\caption{Abundance Results for Galaxies with T$_{e}$ Measurements} \label{tab:elsa} \\

\hline \hline
\noalign{\smallskip}
& & \multicolumn{6}{c}{Abundance Ratios}\\
& & \cline{1-6}
\noalign{\smallskip}
HII Region   & Log(O/H)+12 & He/H                  & N/H                   & O/H		   & Ne/H		   & S/H                   & Ar/H \\
(1)          & (2)         & (3)                   & (4)                   & (5) 		   & (6)  		   & (7) 		   & (8) \\
\noalign{\smallskip}\hline \noalign{\smallskip}
071352+1031\,\emph{b} & 7.880$\,\pm\,$0.019 & 7.88$\,\pm\,$0.26(-2) & 2.87$\,\pm\,$0.17(-6) & 7.59$\,\pm\,$0.33(-5) & 1.47$\,\pm\,$0.07(-5) & 3.21$\,\pm\,$0.14(-7) & 2.77$\,\pm\,$0.13(-7)\\
081821+2431\,\emph{a} & 8.068$\,\pm\,$0.053 & 8.59$\,\pm\,$0.31(-2) & 3.91$\,\pm\,$0.42(-6) & 1.17$\,\pm\,$0.14(-4) & 2.71$\,\pm\,$0.38(-5) & 2.00$\,\pm\,$0.52(-6) & 2.28$\,\pm\,$0.31(-7)\\
115906+2428\,\emph{a} & 8.009$\,\pm\,$0.050 & 8.62$\,\pm\,$0.48(-2) & 4.09$\,\pm\,$0.42(-6) & 1.02$\,\pm\,$0.12(-4) & 2.06$\,\pm\,$0.27(-5) & 1.88$\,\pm\,$0.27(-6) & 4.20$\,\pm\,$0.48(-7)\\
115906+2428\,\emph{b} & 7.851$\,\pm\,$0.053 & 7.39$\,\pm\,$0.76(-2) & 2.82$\,\pm\,$0.37(-6) & 7.10$\,\pm\,$0.86(-5) & 1.65$\,\pm\,$0.24(-5) & 6.17$\,\pm\,$0.61(-7) & 2.77$\,\pm\,$0.40(-7)\\
\end{deluxetable*}
\normalsize


\scriptsize
\begin{deluxetable*}{l c c c c c}

\caption{Global Oxygen Abundance} \label{tab:av}\\

\hline \hline\noalign{\smallskip}
ADBS Galaxy  & Log(O/H)+12 & N$_{\textrm{T$_{e}$}}$ & N$_{\textrm{McG}}$ & Standard Deviation & Range \\
(1)          & (2)         & (3)                   & (4)  & (5) & (6)   \\
\noalign{\smallskip}\hline \noalign{\smallskip}
071352+1031 & 7.89$\,\pm\,$0.02 & 1 & 1 & 0.093 & 7.88 - 8.01 \\
081821+2431 & 8.00$\,\pm\,$0.05 & 1 & 1 & 0.048 & 8.07 - 8.14 \\
100508+2207 & 8.04$\,\pm\,$0.10 & 0 & 1 & --         & --	        \\
100735+1306 & 8.09$\,\pm\,$0.10 & 0 & 1 & --         & --		\\
113845+2008 & 8.71$\,\pm\,$0.10 & 0 & 1 & --         & --		\\
115906+2428 & 7.93$\,\pm\,$0.04 & 2 & 0 & 0.112 & 7.80 - 8.01 \\
121206+2518 & 7.97$\,\pm\,$0.05 & 1 & 0 & --         & --                  \\
125156+1205 & 8.87$\,\pm\,$0.03 & 0 & 10 & 0.067 & 8.78 - 8.99 \\
125850+1308 & 8.05$\,\pm\,$0.03 & 3 & 6 &0.103 & 7.91 - 8.18\\
135822+2533 & 7.97$\,\pm\,$0.02 & 4 & 4 & 0.051 & 7.94 - 8.04\\
145647+0930 & 8.04$\,\pm\,$0.03 & 2 & 4 & 0.071 & 7.93 - 8.15\\
153703+2009 & 7.73$\,\pm\,$0.03 & 2 & 3 & 0.066 & 7.61 - 7.77\\
\noalign{\smallskip}\hline 
\hline
\caption{Global oxygen abundance for each galaxy, based on weighted average of the best abundance measurement available for each \hii\ in this study. Error given in Column 2 corresponds to the error on the weighted mean.  Columns 3 \& 4 indicate the number \hii\ regions for which \te\ or McGaugh abundances (respectively) were used to calculate global average.  Columns 5 \& 6 give the standard deviation about the weighted mean and the complete range spanned by the individual \hii\ regions in each galaxy} \\
\end{deluxetable*}
\normalsize

\footnotesize
\begin{deluxetable*}{c c c c c c }
\caption{R$_{23}$ McGaugh Method Abundance Results and Key Line Ratios} \label{tab:mcg} \\

\hline \hline\noalign{\smallskip}
HII Region    & R$_{23}$ & O$_{23}$ & [\ion{N}{2}]/[\ion{O}{2}] & log(O/H)+12 & $\Delta$ log(O/H)]\\
&  &  &   & & [Mcg - \te]\\
\noalign{\smallskip}\hline \noalign{\smallskip}
071352+1031\,\emph{a} & 0.871 & 0.390 & -1.159 & 8.01 & -0.09	     \\
071352+1031\,\emph{b} & 0.879 & 0.546 & -1.206 & 7.97 & ...	     \\ \noalign{\smallskip}
081821+2431\,\emph{a} & 0.890 & 0.111 & -1.206 & 8.18 & +0.11	     \\
081821+2431\,\emph{b} & 0.815 &-0.181 & -1.172 & 8.14 & ... 	     \\ \noalign{\smallskip}
100508+2207\,\emph{a} & 0.882 & 0.387 & -1.256 & 8.04 & ... 	     \\
\end{deluxetable*} 
\normalsize

\scriptsize
\begin{deluxetable*}{l c c c c}

\caption{Average Abundance Ratios} \label{tab:other}\\

\hline \hline\noalign{\smallskip}
Source  & log(Ar/O) & log(Ne/O) & log(S/O) & log(N/O) \\
\noalign{\smallskip}\hline \noalign{\smallskip}
This Paper      & $-2.49\,\pm\,$0.11 & $-0.68\,\pm\,$0.07 & $-1.67\,\pm\,$0.09 & $-1.39\,\pm\,$0.09\\
\citet{vzh06}    & $-2.19\,\pm\,$0.09 & $-0.79\,\pm\,$0.09 & $-1.53\,\pm\,$0.09 & $-1.41\,\pm\,$0.15\\
\citet{mel04}    & $-2.20\,\pm\,$0.13 & $-0.65\,\pm\,$0.07 & $-1.47\,\pm\,$0.14 & $-1.42\,\pm\,$0.12\\
\citet{it99}         & $-2.26\,\pm\,$0.09 & $-0.72\,\pm\,$0.06 & $-1.56\,\pm\,$0.06 & $-1.47\,\pm\,$0.14\\
\noalign{\smallskip}
ADBS 113845+2008\emph{a}:\\
(T${e}$ $\approx$ 7k K) & $-2.41\,\pm\,$0.11 & $-0.59\,\pm\,$0.14 & $-1.41\,\pm\,$0.30 & $-1.13\,\pm\,$0.15\\
\noalign{\smallskip}\hline 
\hline
\caption{Average relative abundances of $\alpha$-process elements and nitrogen from the literature and this sample.  The last line includes the relative abundance for these elements in ADBS 113845+2008 using an electron temperature derived from the McGaugh method.  The errors for ADBS 113845+2008 were calculated with an error for \te\ of approximately $\pm$2000 K.}
\end{deluxetable*}
\normalsize
\clearpage

\clearpage
\begin{figure}
\centering
\vskip 0.5in
\includegraphics[width=5in]{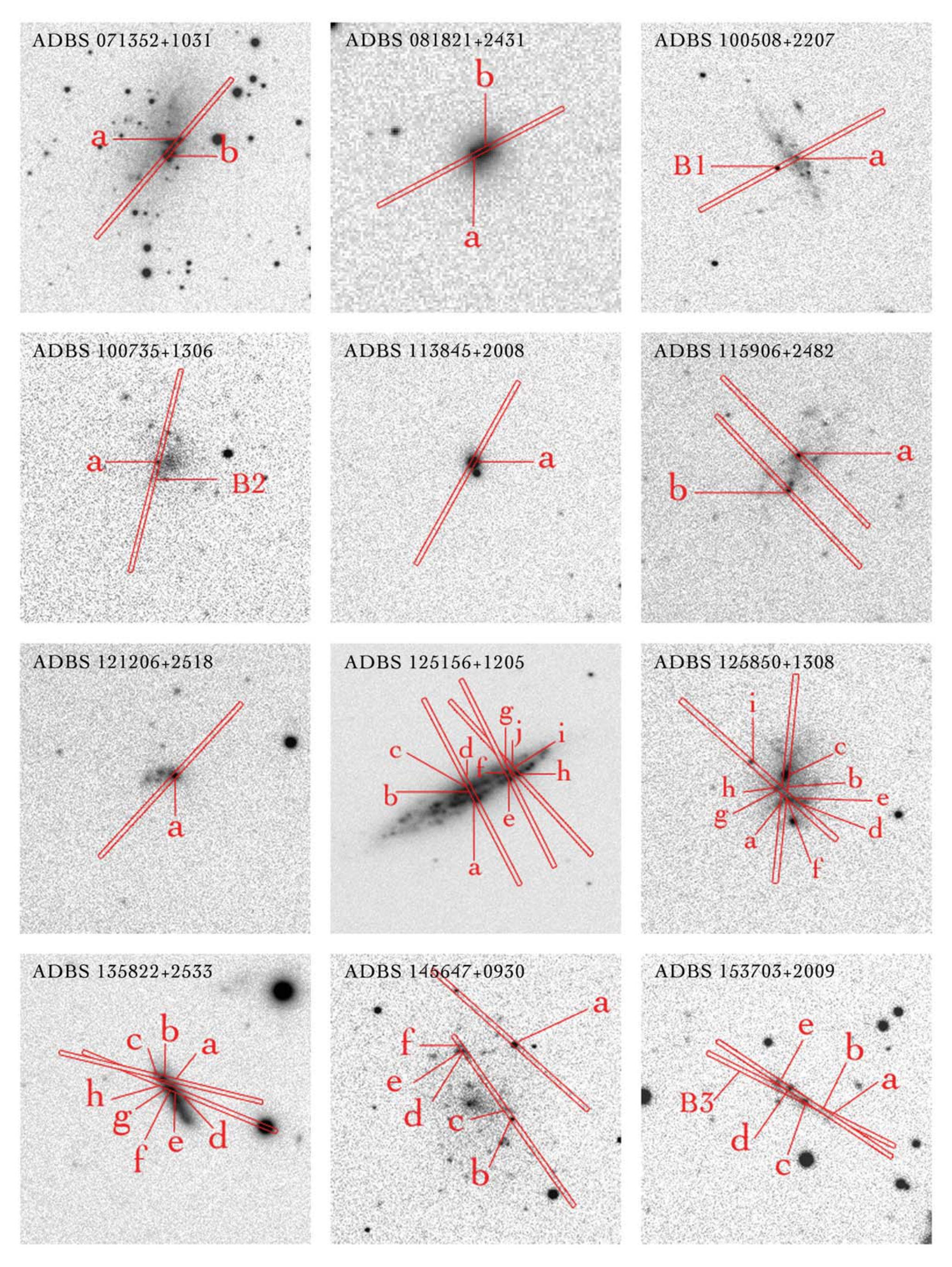}
\vskip -0.1in
\caption{Finder charts for each galaxy discussed in this paper. Images shown here are a 150$\arcsec \times 150\arcsec$ section of the unsubtracted narrowband \ha\ observations. The \hii\ region designations presented here are the same throughout the paper. Regions designated with B\,1-3 are background sources described in Section \ref{subsec:ident}.}
\label{fig:finder}
\end{figure}

\clearpage
\begin{figure}
\centering
\vskip 0.5in
\includegraphics[width=6in]{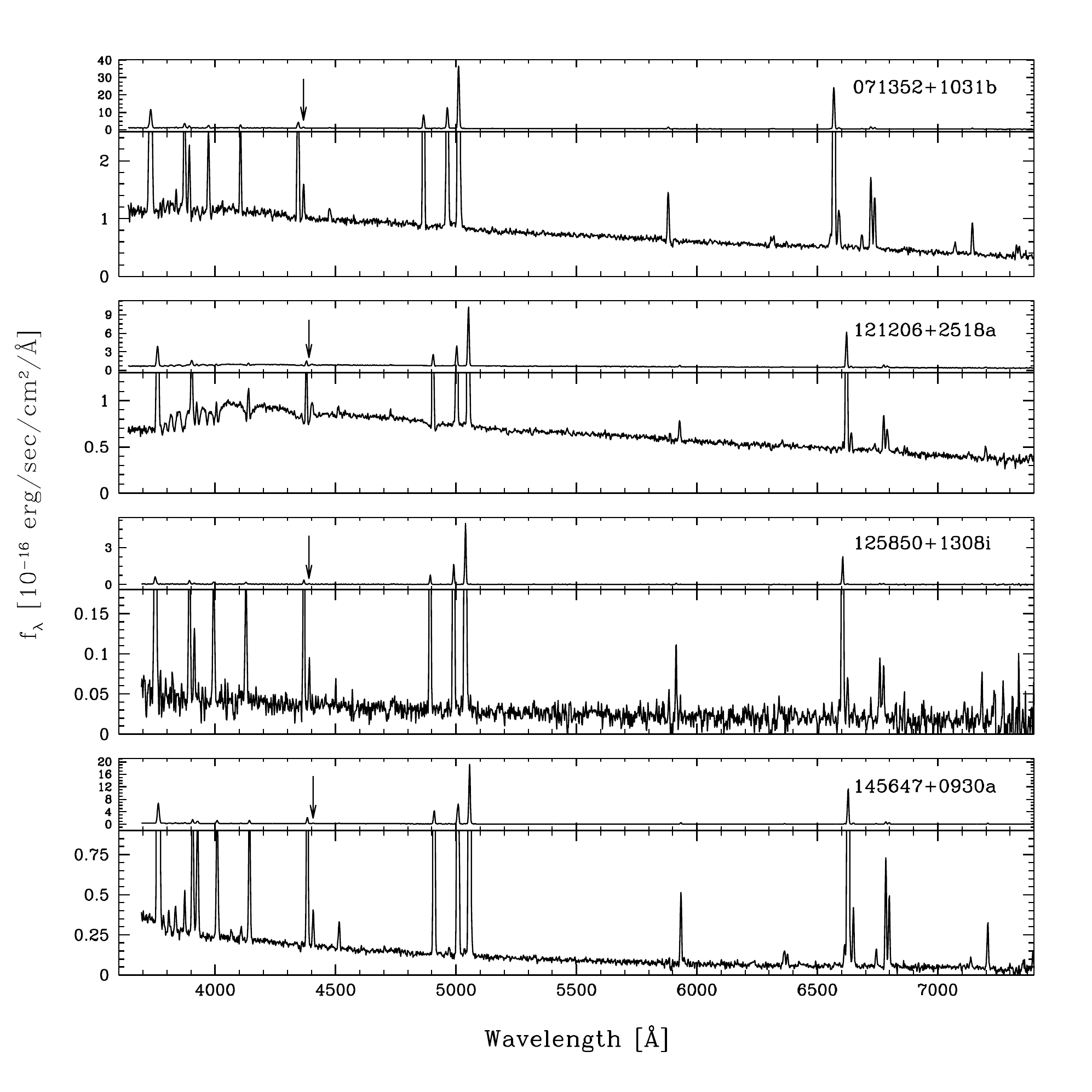}
\vskip -0.1in
\caption{Four examples of high signal-to-noise spectra containing a detectable [\ion{O}{3}]$\lambda$4363 emission line. The location of the [\ion{O}{3}]$\lambda$4363 line is marked with a small arrow.}
\label{fig:specfirst}
\end{figure}

\clearpage
\begin{figure}
\centering
\vskip 0.5in
\includegraphics[width=6in]{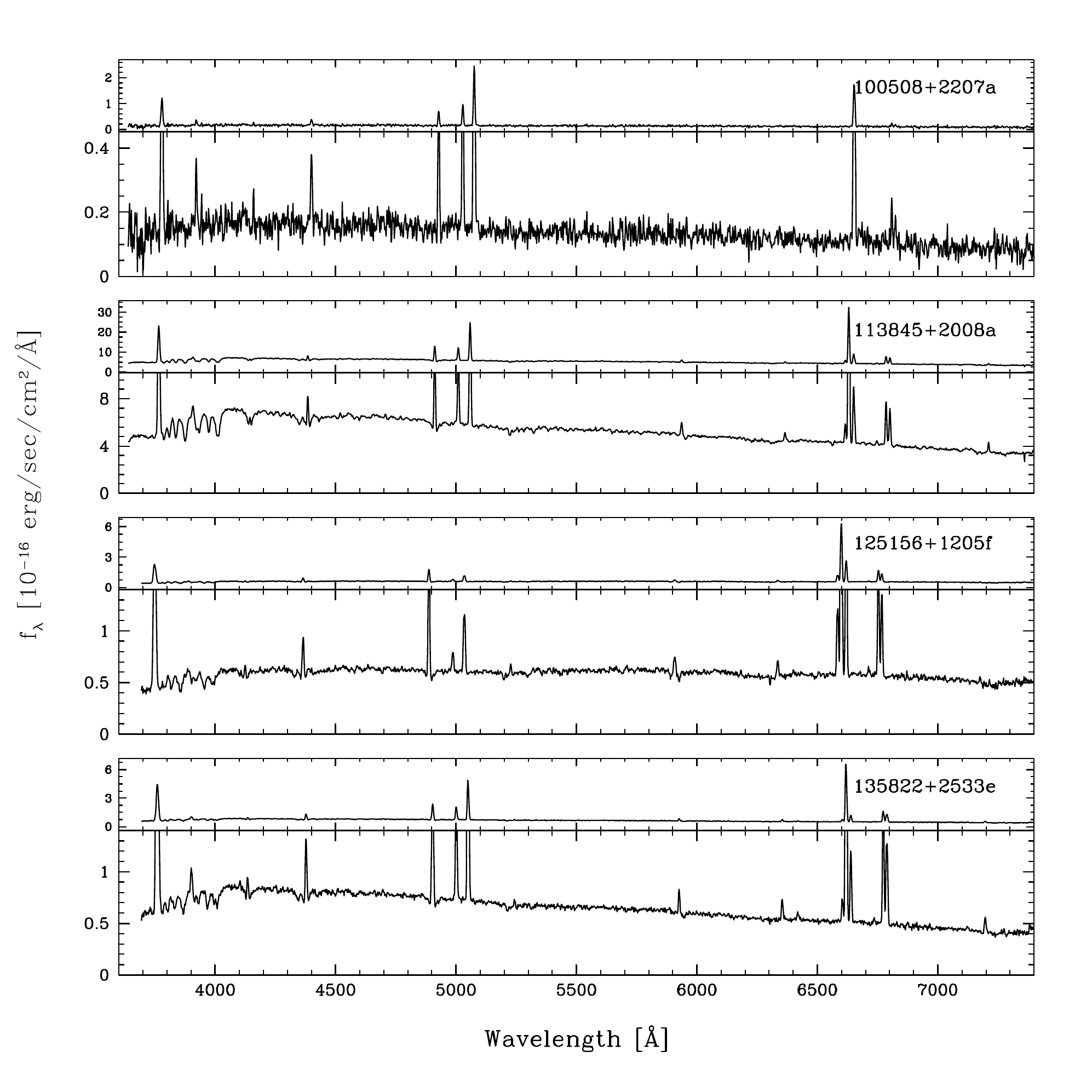}
\vskip -0.1in
\caption{Four example spectra without an [\ion{O}{3}]$\lambda$4363 emission line. These spectra respresent the portion of our sample with lower signal-to-noise or \hii\ regions that where the weak lines were not strong enough to be detected.}
\label{fig:speclast}
\end{figure}

\clearpage
\begin{figure}
\centering
\vskip 0.5in
\includegraphics[width=6in]{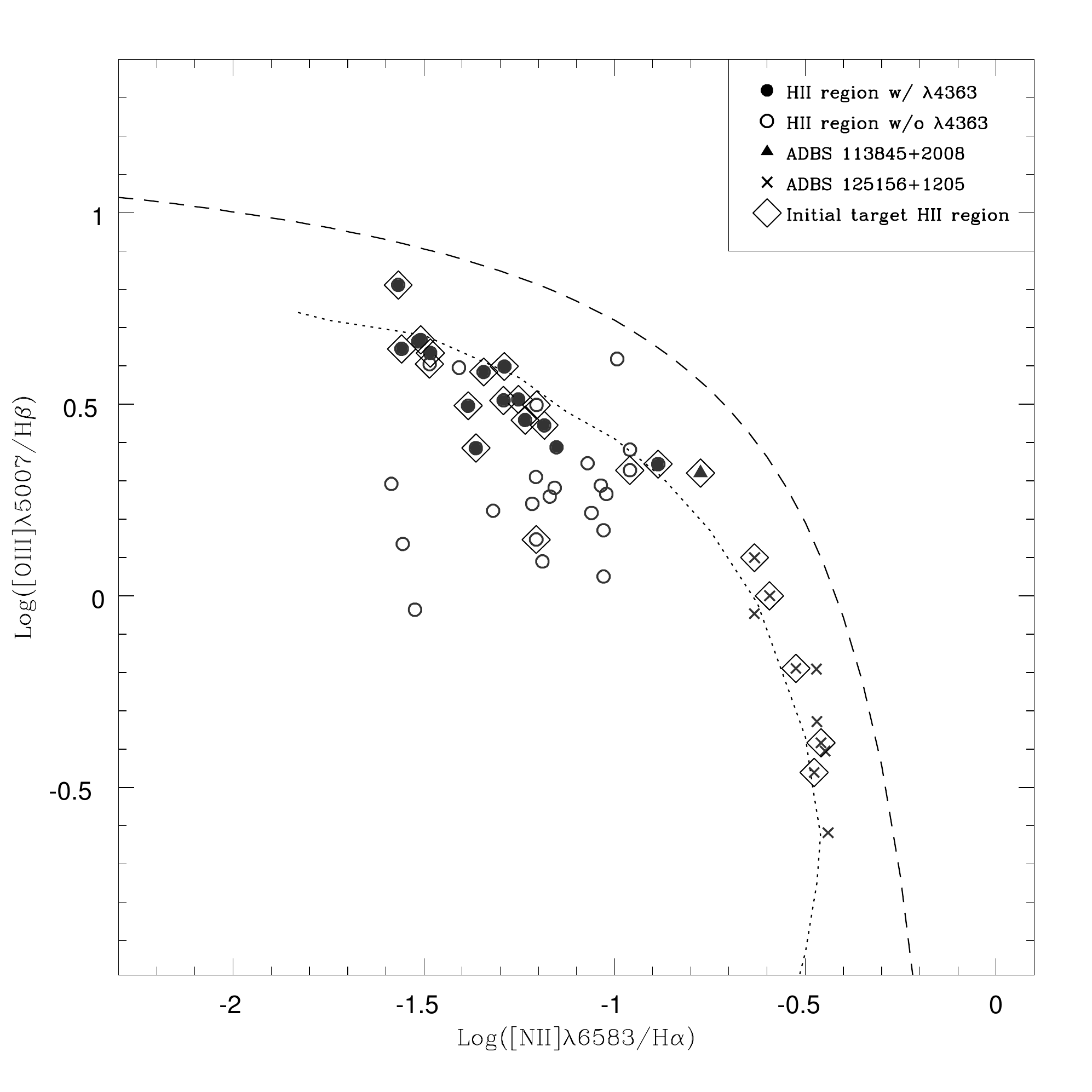}
\vskip -0.1in
\caption{Emission-line diagnostic diagram.  Each \hii-region is plotted separately and the meaning of the different symbols is indicated on the legend in the upper right. Filled points represent \hii-regions with detectable \oiiite, while the open points are \hii-regions were \oiiite\ was not detectable; points enclosed in diamonds represent the target \hii-regions chosen from the \ha\ images.  The \texttt{x}'s are \hii-regions within ADBS 125156+1205 and the triangle is ADBS 113845+2008. The solid line traces the star-forming galaxy models of \citet{de86}; the dashed line \citep{kau03} delineates the region of the diagram usually occupied by AGN.}
\label{fig:diag}
\end{figure}

\clearpage
\begin{figure}
\centering
\vskip 0.5in
\includegraphics[width=6in]{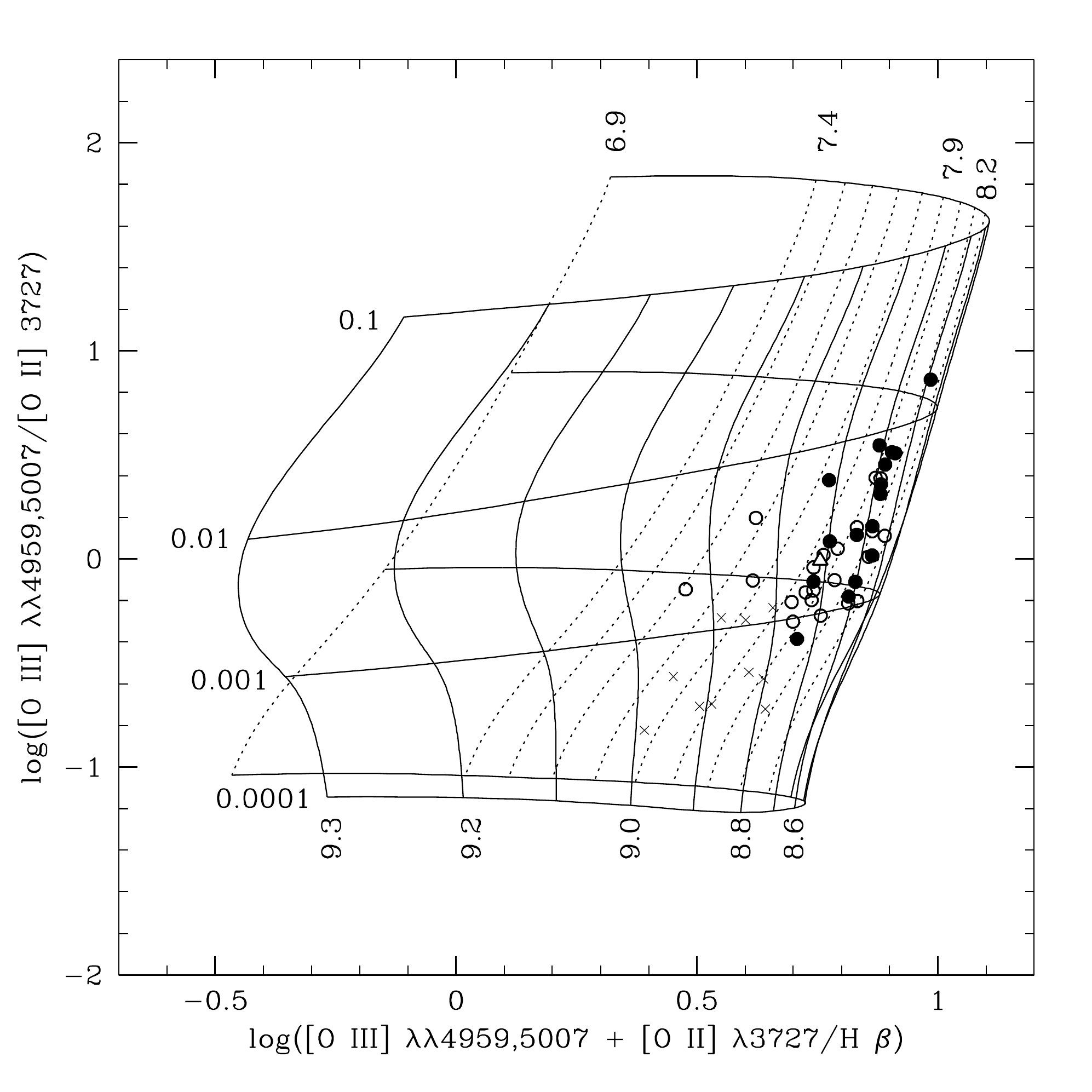}
\vskip -0.1in
\caption{McGaugh theoretical grids \citep{mcg} plotted with the \hii\ regions discussed in this paper.  The roughly horizontal lines in the models indicate different ionization parameters; the roughly vertical lines represent lines of constant oxygen abundance, given in the figure as log(O/H)+12.  From log(O/H)+12 = 7.4 to 9.3 a line of constant abundance is given for every 0.1 dex increase.  Similar to Figure \ref{fig:diag} filled points represent \hii\ regions with measurable \oiiite\ emission, \hii\ regions in ADBS 125156+1205 are represented by \texttt{x}'s, and ADBS 113845+2008 is represented with a triangle.} 
\label{fig:mcg}
\end{figure}

\clearpage
\begin{figure}
\centering
\vskip 0.5in
\includegraphics[width=6in]{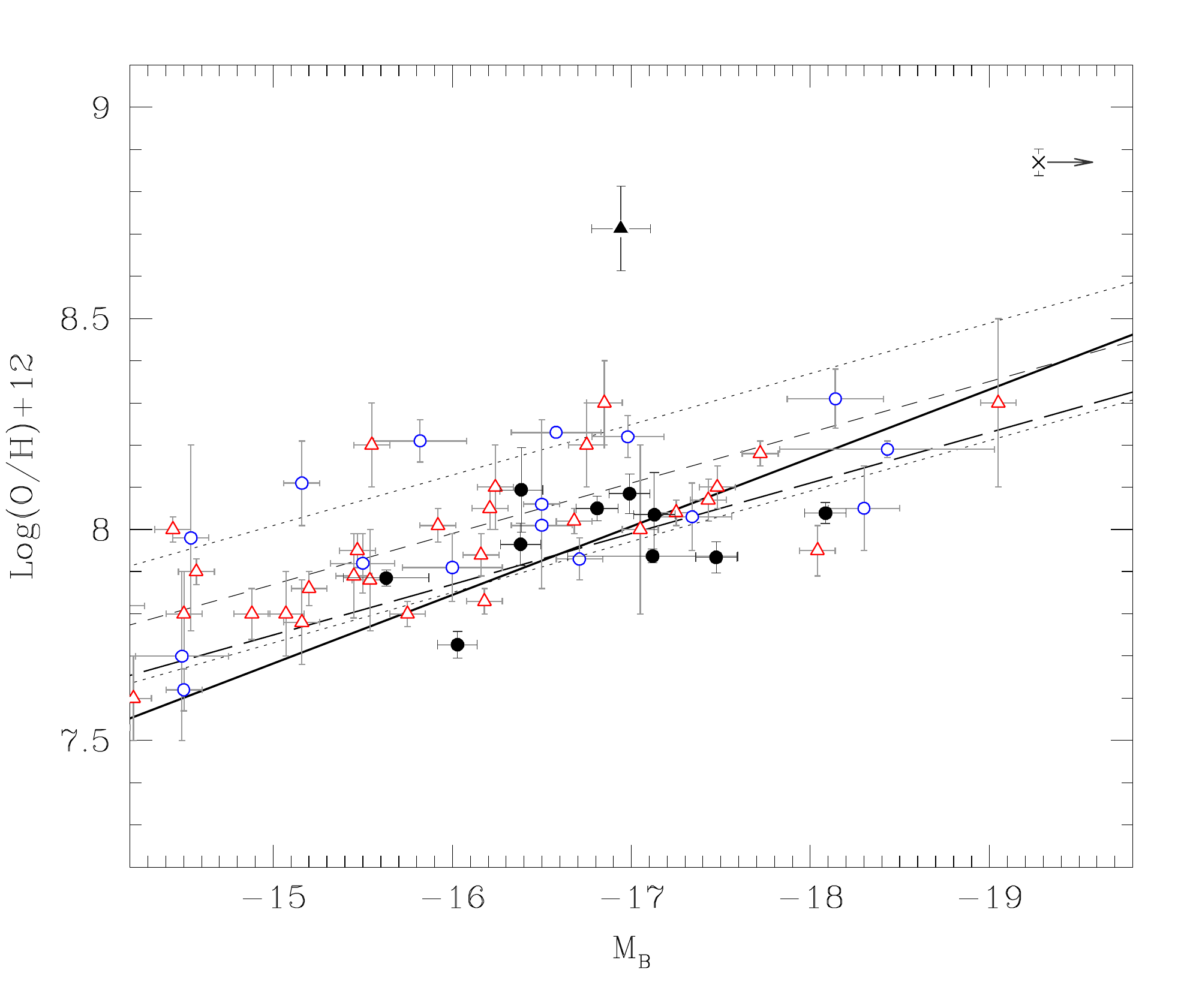}
\vskip -0.1in
\caption{Luminosity-metallicity relationship for this sample (filled black points) as well as dIrr galaxies from the literature (Lee et al. 2006b: open blue triangles, van Zee \& Haynes 2006: red points).  The filled triangle represents ADBS 113845+2008 and the \texttt{x} represents ADBS 125156+1205. The solid bold line is the fit for our sample, excluding ADBS 113845+2008 and ADBS 125156+1205 for the reasons discussed in the text.  The short-dashed line is the fit obtained from fitting the literature sample with the dotted lines above and below representing the 1$\sigma$ error on the intercept. The bold long-dashed line is the fit for our sample holding the slope constant at -0.120 (the slope obtained from fitting the literature sample).}
\label{fig:metlum}
\end{figure} 

\clearpage
\begin{figure}
\centering
\vskip -0.7in
\includegraphics[width=6in]{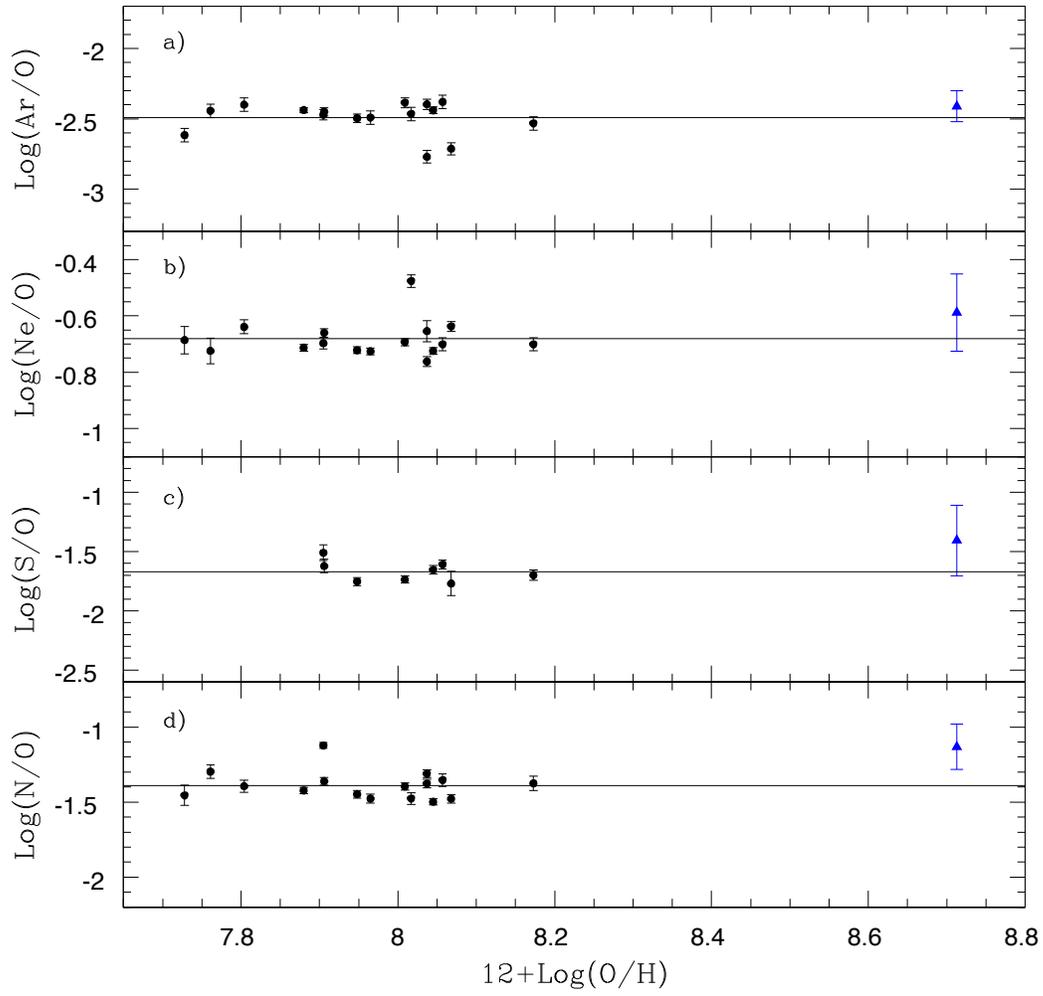}
\vskip -0.1in
\caption{Black points show relative abundances of Ne, S, Ar, and N vs. oxygen abundance for \hii\ regions which have direct temperature determinations (Table \ref{tab:elsa}). Panel c) contains fewer points than the others because an accurate determination of the sulfur abundance was not available unless the \ion{S}{2}$\lambda6312$ was observable and measurable. The dotted line represents the average abundance in each case. The blue triangle represents ADBS 113845+2008\emph{a} where the elemental abundances were calculated using the electron temperature determined from the McGaugh abundance.} 
\label{fig:other}
\end{figure}

\clearpage
\begin{figure}
\centering
\vskip 0.5in
\includegraphics[width=6in]{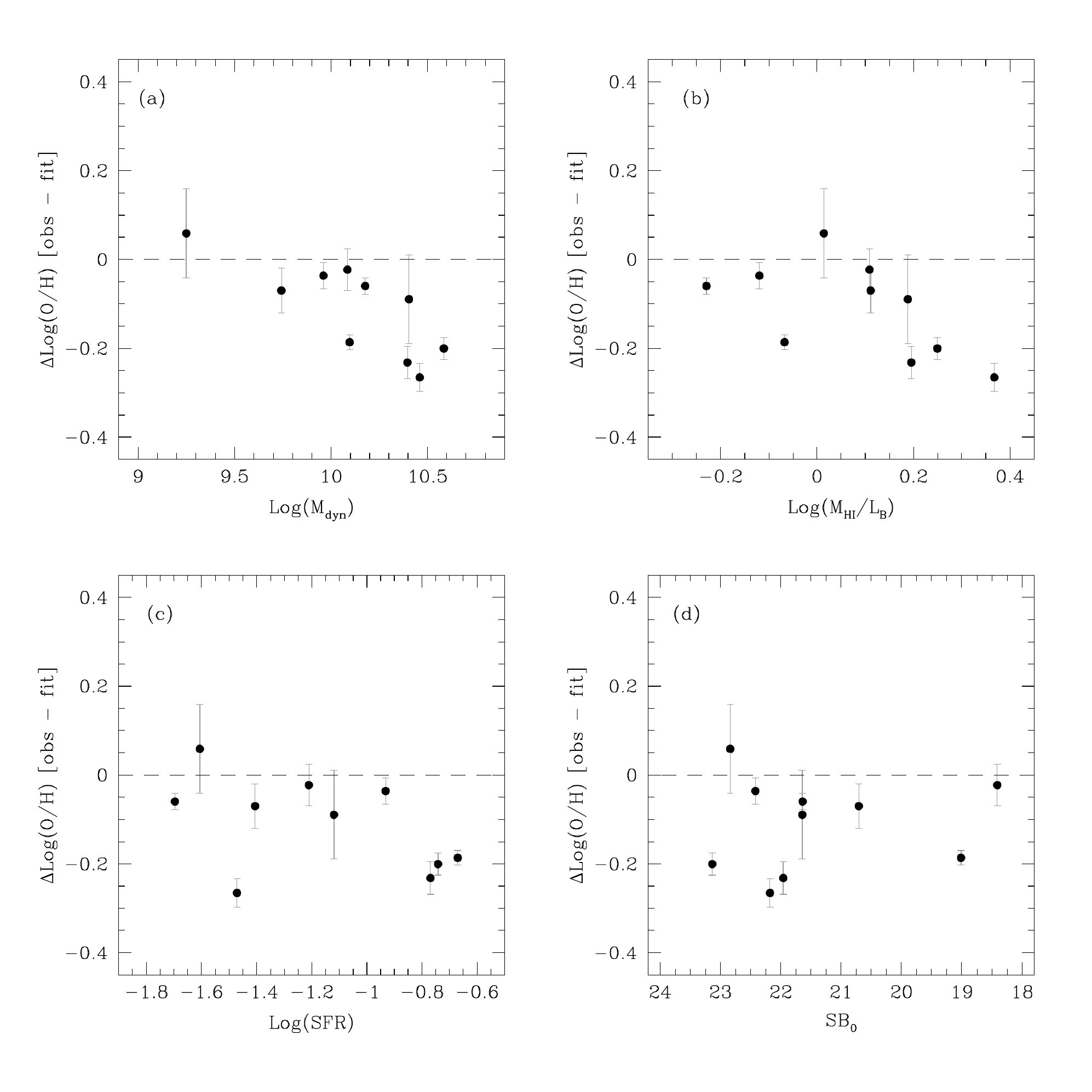}
\vskip -0.1in
\caption{Residual in oxygen abundance (calculated using the fit in eq. (2)) for each of four parameters: a) dynamical mass, b) gas richness, c) star-formation rate, d) central surface brightness.  The dashed line marks the fit to the literature sample and is the same line plotted as a dashed line in Figure \ref{fig:metlum}.}
\label{fig:res_fig}
\end{figure}

\clearpage
\begin{figure}
\centering
\vskip 0.5in
\includegraphics[width=6in]{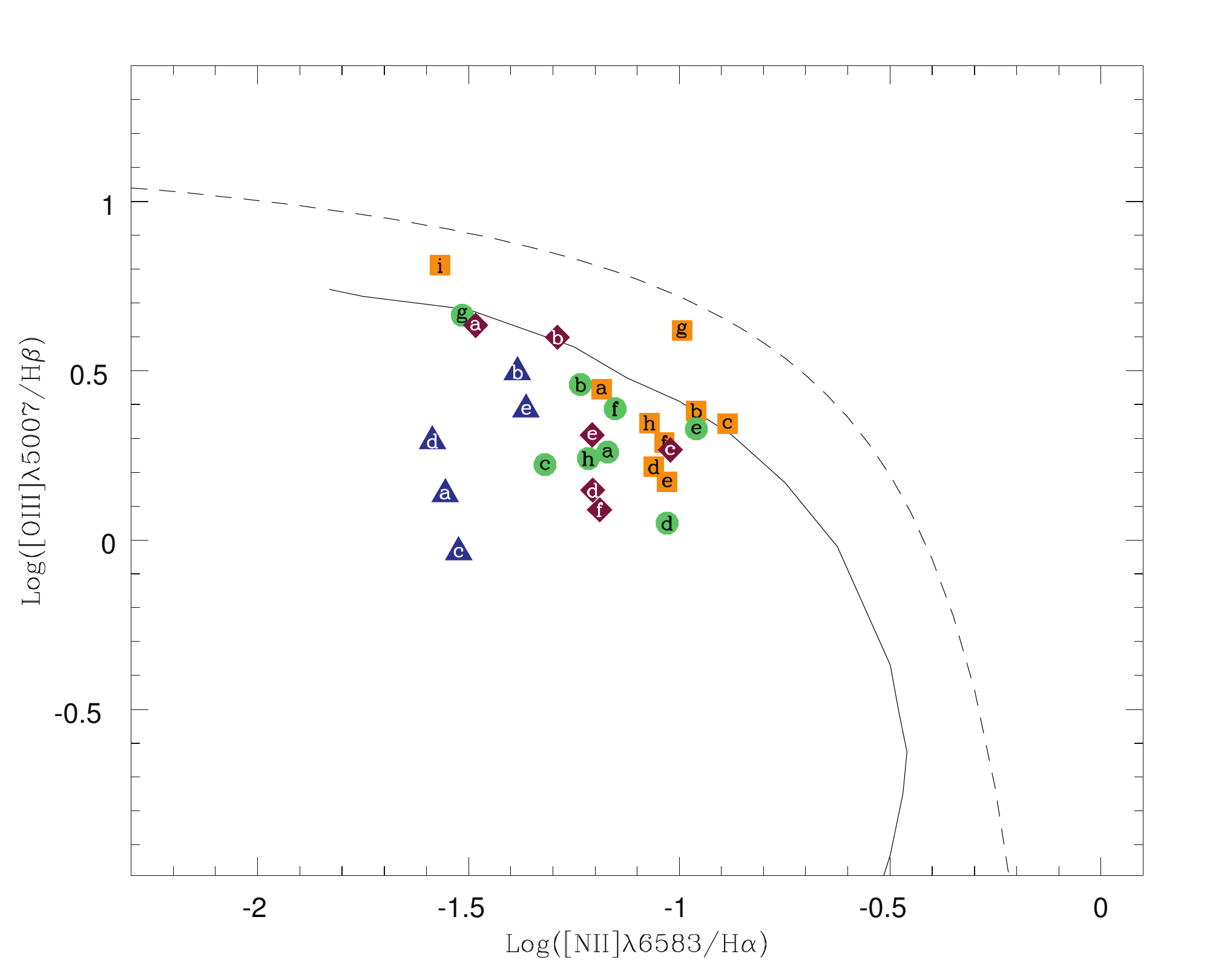}
\vskip -0.1in
\caption{Diagnostic diagram with \hii\ regions from 4 different galaxies displayed: ADBS 125850+1308 (orange squares), ADBS 135822+2533 (green circles), ADBS 145647+0930 (burgundy diamonds), and ADBS 153703+2009 (blue triangles).  The letter in each symbol indicates which \hii\ region the point represents.}
\label{fig:ind}
\end{figure}

\end{document}